\documentclass{emulateapj}

\usepackage{amsmath}
\usepackage{amssymb}
\usepackage{epsfig}
\usepackage{graphicx}
\usepackage{ifthen}
\usepackage{latexsym}
\usepackage{rotating}
\usepackage{subfigure}
\usepackage{times,epsf}
\usepackage{txfonts}
\usepackage{varioref}
\usepackage{verbatim}
\usepackage{url}
\usepackage{color}
\usepackage[dvipsnames]{xcolor}
\usepackage[T1]{fontenc}
\usepackage{float}
\usepackage[normalem]{ulem}

\begin{document}

\begin{abstract}

We develop a Monte Carlo code to compute the Compton-scattered X-ray flux arising from a hot inner flow that undergoes Lense-Thirring precession. The hot flow intercepts seed photons from an outer truncated thin disk. A fraction of the Comptonized photons will illuminate the disk and the reflected/reprocessed photons will contribute to the observed spectrum. The total spectrum, including disk thermal emission, hot
flow Comptonization, and disk reflection, is modeled within the framework of general relativity, taking light bending and gravitational redshift into account. The simulations are performed in the context of the Lense-Thirring precession model for the low-frequency quasi-periodic oscillations, so the inner flow is assumed to precess, leading to periodic modulation of the emitted radiation.

In this work, we concentrate on the energy-dependent X-ray variability of the model and, in particular, on the evolution of the variability during the spectral transition from hard to soft state, which is implemented by the decrease of the truncation radius of the outer disk towards Innermost Stable Circular Orbit (ISCO). In the hard state, where the Comptonizing flow is geometrically thick, the Comptonization is weakly variable with a fractional variability amplitude of $\leq$10\%; in the soft state, where the Comptonizing flow is cooled down and thus becomes geometrically thin, and the fractional variability of the Comptonization
is highly variable, increasing with photon energy. The fractional variability of the reflection increases with energy, and the reflection emission for low spin is counterintuitively more variable than the one for high
spin.

\end{abstract}


\title{X-ray quasi-periodic oscillations in Lense--Thirring precession model - I.  variability of relativistic continuum}

\author{Bei You\altaffilmark{1,2}, Michal Bursa\altaffilmark{3} and Piotr T. \.{Z}ycki\altaffilmark{2}}
\altaffiltext{1}{School of Physics and Technology, Wuhan University, Wuhan 430072, China; youbei@whu.edu.cn}
\altaffiltext{2}{Nicolaus Copernicus Astronomical Center, Polish Academy of Sciences, Bartycka 18, 00-716 Warsaw, Poland}
\altaffiltext{3}{Astronomical Institute, Academy of Sciences, Bo\v cn\'i II 1401, 14131 Prague, Czech Republic}

\maketitle

\section{Introduction}


In black hole X-ray binaries, matter from the companion stars is accreted onto the black hole through stellar wind or Roche lobe, forming the accretion disk.
The viscosity in the disk is thought to play an important role in transferring the angular momentum outward, so that the flow can spiral in toward the black hole and is heated up at the expense of gravitational energy.

One of the observational appearances of the accretion process is the spectral energy distribution of the emitted X-rays. It is puzzling that even at a similar bolometric luminosity, a given X-ray binary can show significant difference in the spectral shapes with spectral index $\Gamma < 2$ and $\Gamma >2$, leading to the definition of the hard state and the soft state, respectively. The hard state is characterized by the high-energy Comptonization component (peaking at E $\ge$ 100 keV), which dominates over the weak, low-energy blackbody component, whereas the soft state is characterized by a strong (disk) blackbody component below $\sim$ 10 keV and a weak, high-energy tail of $\sim 25\%$ of the total bolometric luminosity (see, e.g., Gierli{\'n}ski et al. 1999; Zdziarski \& Gierli{\'n}ski 2004; Remillard \& McClintock 2006; Done, Gierli{\'n}ski, \& Kubota 2007, for the detailed review). The hard-soft spectral transitions in black hole binary outbursts have been observed with a number of X-ray satellite 
observatories (review in, e.g., Fender \& Belloni 2004), indicating a change in the relative contribution of the spectral components to the overall radiation. Such spectral transition can be interpreted by the truncated disk model (Meyer \& Meyer-Hofmeister 1994; Esin et al. 1997; Liu et al. 1999; R{\'o}{\.z}a{\'n}ska \& Czerny 2000; Qiao \& Liu 2012). The blackbody component is due to the thermal radiation from the geometrically thin, optically thick disk (Shakura \& Sunyaev 1973), which is truncated at a certain radius $R_{\rm tr}$ exceeding ISCO ($R_{\rm ISCO} = 6$ for $a=0$; radius is in units of the gravitational radius throughout this work), whereas the Comptonization emission arises from the hot flow interior to the disk. In the hard state, the truncation radius is tens of gravitational radii away from the black hole, and thus the disk dissipates a small fraction of gravitational energy, while most of the energy is radiated from the inner flow, as the dominant Comptonization component at high energy. As the source brightens, the truncation radius moves inward, which will increase the blackbody radiation luminosity. The hot flow will then be cooled down owing to the increased flux of soft photons from the disk, and consequently  the hard X-ray tail is softened.  
The truncated disk model is supported by the observations of the disk receding in the soft-hard transition (Gierli{\'n}ski et al. 2008).   

Another observational appearance of the accretion process is the broadband X-ray variability, covering timescales from from $\le 0.01$ 
s to thousands of seconds (Yu et al.2003; Belloni \& Motta 2016). Fourier analysis and other techniques have been used to study the variability property. The derived power spectral density (PSD) is generally described by a broadband noise continuum with narrow peaks superimposed on this. The narrow peaks indicate quasi-periodic oscillations (QPOs) within the flow. However, the physical origin of QPOs is still under debate. As for low-frequency (1$\sim$10 Hz) quasi-periodic oscillations (LFQPOs), the possible mechanism involves the so-called Lense-Thirring precession which is due to the frame-dragging effect in the strong gravity field (see Stella \& Vietri 1998; Stella et al. 1999; see Belloni \& Motta 2016, for the reviews of other alternative QPO models). In the truncation geometry, if the flow interior to the disk misaligns with black hole spin, it will precess around spin axis at a given frequency. 
Since the hot flow is thought to produce hard X-ray emission via inverse Comptonization, the Lense-Thirring precession of the hot flow will result in the X-ray modulation. Motivated by this scenario, Ingram et al. (2009) incorporated the truncated model with Lense-Thirring precession to explain the observed low frequency QPOs in the X-ray variability.

Besides matching the predicted QPO frequency in the Lense-Thirring precession model with the observation data, it is necessary to verify whether this precession scenario of inner flow can reproduce the observed fractional variability amplitude, which is defined as the ratio of root-mean-squared (rms) variability over mean flux. The energy-dependent fractional rms, i.e., fractional rms spectrum, was observationally explored in the literature (e.g., Rodriguez et al. 2002,2004). The energy-resolved variability enables one to decompose the contribution from different variable emission components (e.g., the disk blackbody, the Comptonization of the hot plasma, and the reflection off the disk) and hence determine how the spectral components vary with respect to each other (e.g., Sobolewska \& {\.Z}ycki 2006; Axelsson et al. 2014; Axelsson \& Done 2016, in which the fractional rms spectra were studied only for the QPOs.
it is also possible to study the fractional rms over broadband Fourier frequency (Gierli{\'n}ski \& Zdziarski 2005; De Marco et al. 2015), but here we concentrate only on the fractional rms associated with QPOs.



Furthermore, the fractional rms spectrum changes dramatically during the spectral transition of the binary outburst (Rodriguez et al. 2004; Sobolewska \& {\.Z}ycki 2006; Gierlinski et al. 2010). Observationally, within the energy range of $3-30$ keV, the fractional rms spectrum is constant across photon energy in the hard state, while slightly decreasing with energy in the intermediate state and increasing with energy in the soft state (Gierli{\'n}ski \& Zdziarski 2005). In Sobolewska \& {\.Z}ycki (2006), it was shown that the rms spectrum is harder than the time-averaged spectrum in the soft spectral state, while the rms spectrum is softer than the time-averaged spectrum when the source is in the hard spectral state, for some binary systems. 

Therefore, the acceptable model should be the one that can not only explain the origin of the variability but also, more crucially, self-consistently associate the changing of the variability pattern with the varying spectral components during the state transition of the source.
In this paper, we will investigate the evolution of fractional rms spectra during the spectral transition from hard to soft state in the Lense-Thirring precession model of the truncated disk/hot inner flow.

Fundamentally, since the varying X-ray radiation comes from the region where is in the strong gravitational field region, it is necessary to take the relativistic effects into account, i.e., Doppler and gravitational redshift, light bending, and the existence of black hole event horizon, which might affect the variability amplitude and the fractional rms spectrum. 
The combination of these relativistic corrections will reshape not only the X-ray energy spectra (You et al. 2012), but also the fractional rms spectra (Schnittman \& Rezzolla 2006; Schnittman et al. 2006). 
 
In this paper, we simulate the X-ray modulation due to the Lense-Thirring precession of the inner flow in the truncated disk geometry, within the framework of general relativity. We aim at investigating the evolution of the fractional rms spectrum as the source softens from the hard state by means of the decrease in the truncation radius. 
Throughout this work, the calculations of the fractional rms are integrated over all Fourier harmonics, i.e. taking the standard deviation of the flux versus QPO phase.
The description of the assumed geometry of the accretion flow, as well as the setup of the Monte Carlo simulation, will be given in Sect. 2. The results of the relativistic X-ray spectra and fractional rms spectra are shown in Sect. 3. The explanation of the evolution of fractional rms spectra during the spectral transition and the influence of the relativistic effect on the variability will be discussed in Sect. 4. The main conclusions in this paper are given in Sect. 5. 

\section{Model}
\subsection{Truncated outer disk and the precessing inner flow}
\begin{figure}
\includegraphics[width=\columnwidth]{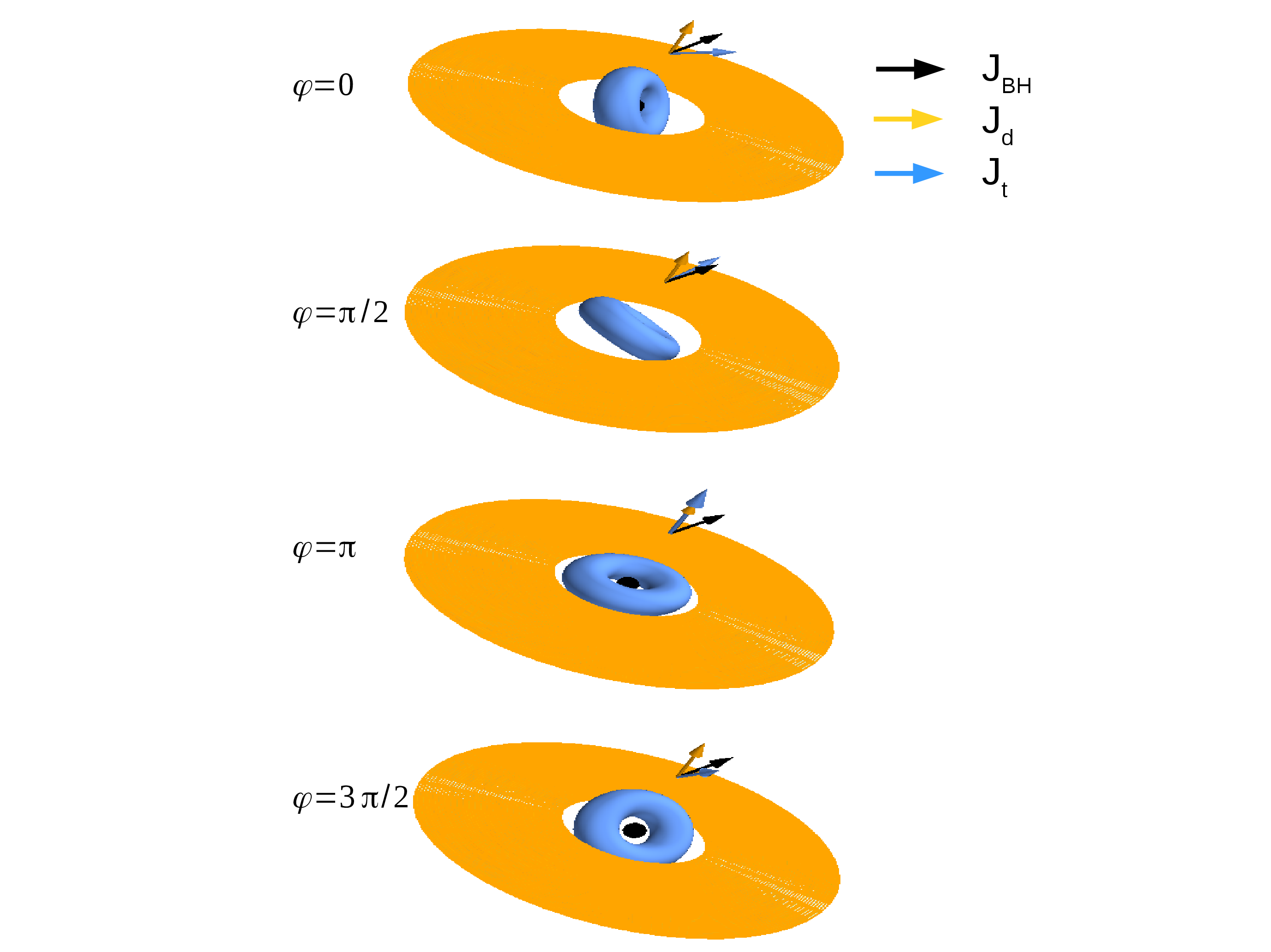}
\caption{
Schematic picture of the entire flow. The accretion flow is assumed to be the truncated at a given radius. Interior to the thin disk is a geometrically thick, optically thin hot flow, which is assumed to have the shape of a torus with a trapezoid cross section. In the context of the Lense-Thirring precession, we assume that the black hole spin axis $J_{\rm BH}$ is tilted with respect to the symmetry axis of the outer disk $J_{\rm d}$ , and the symmetry axis of the torus $J_{\rm t}$ misaligns with the spin axis. Therefore, the precession of the hot flow can be manifested as $J_{\rm t}$ precessing around $J_{\rm BH}$ at some frequency.  
\label{schematic_torus}}
\end{figure}

In this work, the basic geometry of the accretion flow is assumed to be the truncated disk model in which the geometrically thin and optically thick disk 
extends from the outer radius $R_{\rm out}$ down to the truncated radius $R_{\rm tr}$. Interior to the thin disk is a geometrically thick, optically thin hot flow. This is assumed to have the shape of a torus with a trapezoid cross section and it will be referred to as the torus throughout this work. The inner radius of the torus is $R_{\rm in}$, while its outer radius coincides with the inner radius of the disk, $R_{\rm tr}$. The half-opening angle of the torus (measured from the equatorial plane) is $\theta_{\rm o}$. The schematic picture of the entire flow is shown in Fig. \ref{schematic_torus}. In the context of the Lense-Thirring precession, we assume that the black hole spin axis $J_{\rm BH}$ is tilted with respect to the symmetry axis of the outer disk $J_{\rm d}$ by $\theta_{\rm p}$, and the symmetry axis of the torus $J_{\rm t}$ misaligns with the spin axis by $\theta_{\rm t}$. Therefore, the precession of the hot flow can be manifested as $J_{\rm t}$ precessing around $J_{\rm BH}$ at some frequency. 
The geometry used here closely follows works by Ingram \& Done (2012), Veledina et al.(2013), and Ingram et al. (2015).

Note that once the geometrical configuration is fixed, the results of the model calculation will depend on the observer position, which is defined by two angles: the polar angle, $i$, with respect to the symmetry axis of the outer disk $J_{\rm d}$, and the azimuthal angle, $\Phi_{\circ}$, with respect to the plane formed by the black hole spin axis, $J_{\rm BH}$, and the symmetry axis of the outer disk, $J_{\rm d}$. 
For simplicity, most of the results in this paper (e.g., Fig. \ref{spec_0p3_t15}-\ref{rms_10rg_t45} below) will be for the viewer azimuth $\Phi_{\circ} = 0$, i.e.,  the spin axis being directed toward the observer. The impact of the viewer azimuthal angle on the predictions of the variability property will be discussed in Sect. 4.4.

\subsection{Spectral state transitions}
We aim at studying the evolution of spectral and variability properties during the spectral transition from the hard to soft state. In the truncated disk geometry, for given black hole mass and accretion rate, the spectral transition can be implemented by changing the truncation radius (e.g. Esin et al.\ 1997). And this will be done by the use of the \textit{eqpair} code\footnote{http://www.astro.yale.edu/coppi/eqpair/} in our work.

Code \textit{eqpair} computes the electron energy distribution resulting from a balance between heating and direct acceleration of particles on one hand and cooling processes on the other. The processes considered include electron-positron pair balance, bremsstrahlung, and Compton cooling, including external soft-photon input. The final electron distribution can be hybrid and thermal/nonthermal, i.e. consisting of a Maxwellian and a power-law part. In this work we assume the purely thermal case, and the code computes the resulting plasma temperature and its optical depth. The code is described in detail in Coppi (1999).
In \textit{eqpair}, the ratio of the hard to soft compactness $l_{\rm h}/l_{\rm s}$, where $l_{\rm h,\rm s} = L_{\rm h,\rm s}\sigma_{\rm T}/Rm_{\rm e}c^3$, and the soft-photon compactness $l_{\rm s}$ are the first two model parameters in approximate order of decreasing importance.
 
At a given truncation radius $R_{\rm tr}$, the energy generation in the inner flow and outer disk is
\begin{equation}
 Q_{\rm h} \propto \int_{R_{\rm in}}^{R_{\rm tr}} R f(R) dR,
\end{equation}
and 
\begin{equation}
 Q_{\rm s} \propto \int_{R_{\rm tr}}^{R_{\rm out}} R f(R) dR,
\end{equation}
respectively, where $f(R)$ is the energy dissipation rate per surface. 
It should be noted that in \textit{eqpair} the Comptonizing source is assumed to be spherical, and the seed soft photons are uniformly distributed inside this region.
However, in our work the basic geometry of the flow is the inner torus plus outer truncated disk. For simplicity, let us assume the ratio of the hard to soft compactness $l_{\rm h}/l_{\rm s} \approx Q_{\rm h}/(\eta_0 Q_{\rm s})$, where $\eta_0$ is defined as the fraction of the soft-photon energy from the outer disk being intercepted by the inner hot flow. The explicit estimate of the compactness ratio requires the physical solution of the dissipated energy in the flow, which is beyond the scope of this work.
In principle, $\eta_0$ mainly depends on the size of the hot flow and its relative geometry with respect to the disk. For simplicity, 
we further assume the phase-averaged value of $\eta_0=0.05$ (i.e., irrespective of the truncation radius and precession phase). 
Therefore, as the truncation radius decreases, the heating-to-cooling ratio monotonously decreases as well, so that the inner flow will be effectively cooled and the spectrum softens.

Besides the compactness ratio $l_{\rm h}/l_{\rm s}$, the soft compactness $l_s$ is another important parameter that can affect the spectral properties as well. However, this parameter is hard to compute, since the geometry of the flow is not spherical in this work. Poutanen \& Coppi(1998) gave the dependencies of $l_{\rm s}$ on $R$ as
\begin{equation}
 l_{\rm s}(R_{\rm tr}) = l_{\rm s,0}({R_{\rm tr,0}}/{R_{\rm tr}})^2, \quad\rm for\quad R_{\rm tr} < R_{\rm tr,0}
\end{equation}
where $l_{\rm s,0}$ is the soft compactness at $R_{\rm tr,0}$.
Here we adopt the scaling relations above to compute the $l_{\rm s}$, by assuming $l_{\rm s,0} = 1$ at $R_{\rm tr,0} = 90$. 

Overall, under the assumptions above, both $l_{\rm h}/l_{\rm s}$ and $l_{\rm s}$ have the monotonous relations with the truncation radius. This will allow us to use the \textit{eqpair} code to simply compute the parameters of the Comptonizing plasma (optical depth $\tau_{0}$, electron temperature $T_{\circ}$) for a given $R_{\rm tr}$. And optical depth and electron temperature will be taken as input parameters in our Monte Carlo simulation to compute the X-ray energy spectrum and the associated variability in the following sections.

\subsection{Monte Carlo simulation of X-ray energy spectrum}

The X-ray radiation that will be simulated consists of three components: the thermal emission from the truncated disk, the hard X-ray emission from 
the torus, and the component due to the reprocessing of the hard X-rays from the torus in the outer disk. Both the Comptonization and reflection components contribute to the X-ray variability. 
The Monte Carlo code developed in this work takes the relativistic effects, e.g., light bending and gravitational redshift, into account.  

The relativistic thermal spectrum of thin-disk flow is produced by the Novikov \& Thorne (1973) model. For given black hole mass $M$, Eddington-scaled mass accretion rate $\dot{m} \equiv \dot{M}/\dot{M}_{\rm Edd}$, and the spin $a$, the radial flux profile $F(r)$ can be determined. The corresponding local blackbody temperature, 
$T(r)$, is given by the Stefan-Boltzmann law. 
A photon's emission radius is generated according to the $F(r)$ distribution, 
while its energy is generated from the Planck distribution with the temperature 
$T(r)$. Details of these algorithms are given in the Appendix. The remaining 
information of the disk soft photon is, the azimuth of the emitting position and 4-momentum. The azimuth of the emitting position is randomly generated between 0 and $2\pi$. The directions of photon 4-momentum are specified by two angles in the local frame, i.e., the azimuthal angle and the polar angle with respect to the normal of the disk. The azimuthal angle of the photon's momentum is randomly generated between 0 and $2\pi$, while the polar angle is randomly generated under the condition that the number of photons per second emitted into equal portions of solid angle depends on the polar angle. The implement for randomly generating this polar angle is given in Equations (9)-(11) in the Appendix.
Relativistic ray tracing is applied to the photons emitted from the disk, computed with 
the use of the sim5 library.\footnote{https://github.com/mbursa/sim5
}
Emitted disk photons will (i) travel to infinity (forming the observed disk spectrum), (ii) or enter the black hole event horizon (being discarded), or (iii) enter the hot flow (being upscattered by the thermal electrons).

If the disk photon enters into the hot flow, it may be Compton scattered by the hot electrons. 
For simplicity, we assume that the entire hot flow rotates in the radius-dependent Keplerian motion and ignore the radial velocity, 
although this might not be the case (Czerny \& You 2016).
The additional component of the flow velocity, due to precession motion, is not taken into account. Before modeling the Comptonization, 
the photon momentum is converted from the Boyer-Lindquist frame to the local rest frame of the torus, following Kulkarni et al. (2011). In the rest frame of the torus, the inverse Compton scattering of soft photons in the hot plasma is simulated using the standard prescriptions of Pozdnyakov et al.(1983) and Gorecki \& Wilczewski (1984; see also Janiuk et al. 2000). The code assumes uniform plasma density and temperature.
At a given $R_{\rm tr}$, the plasma density can be estimated based on the derived optical depth $\tau_{0}$ in Sect. 2.2, while the electron temperature will depend on the position of the torus during the precession. It should be noted that the derived $\tau_{0}$ and $T_{\circ}$ in Sect. 2.2 are averaged values over precession phase.
In principle, the ratio of heating to cooling rate is associated with the ratio of the disk luminosity intercepted by the torus to the total disk emitted luminosity, 
$\eta(\phi)$, which is variable as a function of precession phase. As the torus precesses, it is expected that the heating-to-cooling ratio correspondingly varies during precession phase, leading to the variation of the electron temperatures.
Therefore, in order to reflect this geometrical effect of the precession on the temperature, we estimate the electron temperature $T_e(\phi)$ at each precession phase based on Equations (13) and (14) in Beloborodov (1999), for given optical depth $\tau_{\circ}$, imposing that the phase-averaged temperature matches with the derived temperature $T_{\circ}$. 

Relativistic ray tracing is always taken into account throughout the consecutive inverse Compton scatterings. The Comptonized photon leaving the torus can (i) be lost in the 
black hole; (ii) go to the infinity, forming the observed Comptonization spectrum; and (iii) enter the outer disk, being reprocessed.
 
The reprocessed photons will produce the reflection spectrum including the iron emission line. 
In this paper, for simplicity, we assume that the outer disk is neutral. The reflection code in this simulation is written following prescriptions by George \& Fabian (1991) and \.{Z}ycki \& Czerny (1994). Once the photon emerges away from the disk, ray tracing is applied again, 
including the Doppler shift due to the Keplerian rotation of the accretion disk. In principle, the reflected photon could enter into the torus again. For simplicity, we assume that reflected photons can travel either to the black hole event horizon (being discarded) or to infinity (forming the observed reflection spectrum).

The overall spectrum is modeled in the range of $10^{-3}$ to $10^3$ keV. 
The uncertainties are assigned assuming that the modeled photon counts within each energy bin follow a Poisson distribution.
Throughout the simulations in this work, at energies $E>100$ keV the photon number count is about 400, so that the corresponding uncertainty is relatively high, $\sim 5\%$. In some cases, this leads to unconstrained values for the fractional variability amplitude, as the Monte Carlo noise becomes significant. Therefore, we limit the X-ray variability analysis to $E\leq 100$ keV.

\subsection{Modulation of the X-Ray flux}

Following the procedure above, we can model the X-ray energy spectrum for different geometrical configurations of the torus. The Lense-Thirring precession of the hot flow is thought to be the mechanism producing low-frequency X-ray modulations. Here we focus on the study of X-ray spectral variability as produced solely by these modulations. We compute a sequence of spectra corresponding to consecutive values of the precession phase and calculate the fractional rms amplitude of variability (van der Klis 1989; Vaughan et al. 2003) as a function of energy.
In the following sections, the spectral and variability properties will be studied in detail. For simplicity, the time lags and relativistic time dilution are not taken into account. This is justified, as the precession timescale, $\sim 1\rm s$, is much longer than the photon traveling time scale of roughly milliseconds for X-ray binaries, provided that the size of the X-ray radiation region is about tens of gravitational radii (Basak \& Zdziarski 2016).

Therefore, at a given $R_{\rm tr}$ the modeling strategy is as follows:
\begin{enumerate}
\item Computing $l_{\rm h}/l_{\rm s}$ and $l_{\rm s}$ to derive the electron temperature and Thomson optical depth.
\item At given precession phase (so that the relative position of the torus with respect to the outer disk is fixed), the overall spectrum is modeled with our Monte Carlo simulation.
\item Modeling the spectrum over one complete precession of the flow to derive the X-ray variability. 
\end{enumerate}
We then repeat the three steps above for different truncation radii, 
to derive the evolution of both spectral and variability properties during 
the state transitions.

The inner radius of the hot flow is rather uncertain. Magnetohydrodynamic (MHD) simulations of the precession flow indicate that the inner radius of the tilted geometrically thick flow will increase as the spin increases (Fragile et al. 2009), which is counterintuitive in the sense that the ISCO of a standard disk decreases with the black hole spin. This is due to the additional torque from the misalignment of the flow, which causes the surface density of the inner flow to dramatically decrease in the inner radii. The analytic estimate for the inner radius in this situation is given in Lubow et al.(2002), i.e., $r_i = 3.0(h/r)^{-4/5}a^{2/5}$, which is roughly consistent with the simulation results by Fragile et al.(2009). This description of the inner radius is used in Ingram (2009) to explain the independence of the low-frequency QPO on the spin. Here we also use this description so that the inner radius is approximately $R_{\rm in} = 5, 7,$ and 8 for $a = 0.3, 0.6,$ and 0.9, respectively.
 

\section{Results}
Throughout this work, we assume $M = 10M_{\sun}$, Eddington-scaled mass accretion rate $\dot{m} \equiv \dot{M}/\dot{M}_{\rm Edd} = 0.01$, and misalignment angle of the black hole spin with respect to the the symmetry axis of the disk $\theta_{\rm p} = 15^{\circ}$, while the symmetry axis of the torus with respect to the spin is tilted by $\theta_{\rm t} = 15^{\circ}$, so that at some particular precession phases the torus will align with the outer disk. 
Regarding the assumed misalignment angle, on one hand, the misalignment needs to be large enough to give an observable effect; on the other hand, we may not expect systems with very large misalignments to be common, since a very large one would likely disrupt the binary system. Therefore, systems that survive to become X-ray binaries may be expected to have predominantly modest misalignments (e.g. Fragos et al. 2010). A misalignment of $\sim$10$^\circ$ (Veledina et al. 2013; Ingram et al. 2015) to $\sim$15$^\circ$ (Ingram \& Done 2012) therefore seems fairly reasonable.

\begin{figure}
\includegraphics[width=\columnwidth]{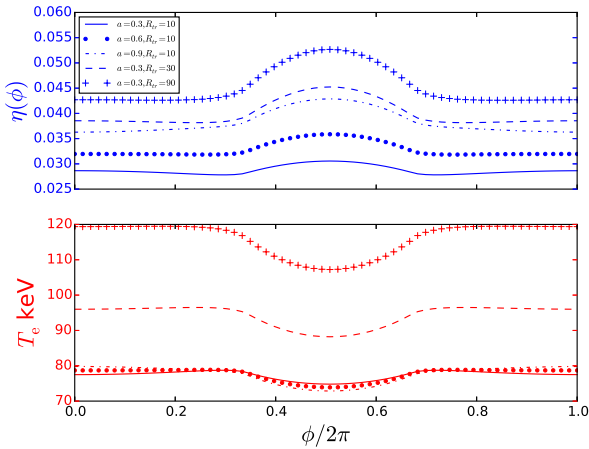}
\caption{
Ratio of the disk luminosity intercepted by the inner hot flow to the total 
disk emitted luminosity, $\eta(\phi)$ (upper panel, in blue) and the resulting electron temperature (bottom panel, in red), as a function
of precession phase, for different black hole spins and disk truncation radii. The solid, dotted, and dotted-dashed lines are for black hole spin $a = 0.3, 0.6,$ and 0.9, respectively, while the truncation radius is fixed at $R_{\rm tr} = 10$. The dashed line and crossed points are for $R_{\rm tr} = 30$ and 90, while the spin is fixed at $a = 0.3$.
The changing relative geometry of the precession flow
leads to the variations of $\eta(\phi)$. 
\label{lsoft}}
\end{figure}

In Fig. \ref{lsoft}, $\eta(\phi)$ and the resultant temperature $T_e(\phi)$ as a function of precession phase, for $R_{\rm tr} = 10, 30,$ and 90,  are plotted, where $a=0.3$.
Additionally, the dependence of $\eta(\phi)$ on the spin is shown.
As the spin increases, the light-bending effect tends to focus many more disk photons toward the torus, resulting in the increase of $\eta(\phi)$.

Based on the geometrical configurations (i.e., relative geometry between the inner torus and outer truncated disk) and physical parameters (i.e., black hole spin, electron Thomson optical depth, and temperature), we are able to compute the emergent spectrum consisting of a disk, Comptonization, and reflection component as a function of the precession phase. The spectral property and energy-dependent fractional variability during the state transition will be analyzed. In the following sections, the results are divided into two classes,  i.e., the half-opening angle $\theta_o = 15^{\circ}$ and $45^{\circ}$, since the realistic value is unknown. 

\subsection{Slim-torus case}
Here we assume a half-opening angle of the torus of $\theta_{\rm o} = 15^{\circ}$.
\subsubsection{Effects of varying the truncation radius}
\begin{figure*}
\includegraphics[width=2\columnwidth]{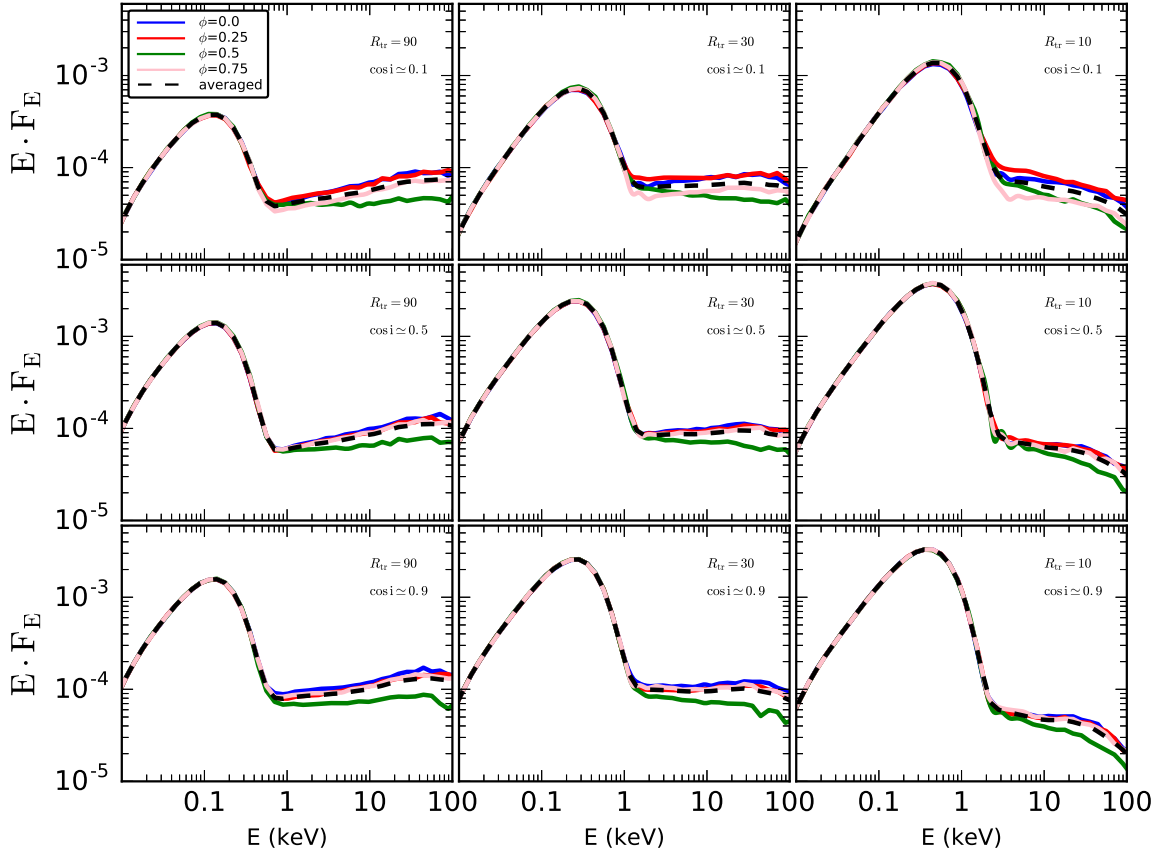}
\caption{
Normalized radiation spectrum from the hot inner flow plus outer truncated disk for the half-opening angle of the flow $\theta_{o} = 15^{\circ}$ and the black hole spin $a=0.3$. The left, middle, and right panels are for the truncation radius $R_{\rm tr} = 90, 30,$ and 10, respectively. The top, middle, and bottom panels are for the sinusoidal value of viewing angle $\cos i$ = 0.1, 0.5, and 0.9, respectively. In each subplot, the spectra at four particular precession phases, which are separated by one-fourth of the precession period, are plotted. The phase-averaged spectrum is plotted in black.
\label{spec_0p3_t15}}
\end{figure*}
In this subcase, we investigate the spectral and variability properties during the state transition from hard to soft state, by adopting the truncation radii $R_{\rm tr} = 90, 30,$ and 10, while black hole spin is fixed at $a = 0.3$. 
In Fig. \ref{spec_0p3_t15}, the overall spectra are plotted, in which the first, second, and third columns are for $R_{\rm tr} = 90, 30,$ and 10, respectively. The top, middle, and bottom panels are for $\cos i = 0.1, 0.5$, and 0.9, respectively, where $i$ is the viewing angle with respect to the symmetry axis of the outer disk.
In each individual subplot, the spectrum consists of the thermal disk emission in the soft energy band, the Comptonization emission from the torus, and the reflection emission off the disk. 
The spectra in four colors correspond to the four particular precession phases separated by $90^{\circ}$ during the precession. 
For the large truncation radius, e.g., $R_{\rm tr} = 90$, a large fraction of the gravitational energy is dissipated in the torus, and the heating-to-cooling ratio is high, so that the electron temperature is high, which leads to a hard spectrum. The disk blackbody radiation peaks at low energy, since the disk is relatively far away from the black hole. As the truncation radius $R_{\rm tr}$ decreases (comparing left panels with right panels), e.g., from $R_{\rm tr} = 30$ to 10, more energy is dissipated in the disk, and the disk luminosity increases. The heating-to-cooling ratio is low, so that the electron will be cooled, producing a soft X-ray spectrum at high photon energy. Therefore, as predicted in the truncated disk model, the X-ray continuum spectrum softens as $R_{\rm tr}$ decreases.
Furthermore, it can be seen that it is the Comptonization rather than the disk emission that varies, by comparing the individual spectrum at different precession phases with the phase-averaged spectrum (black dashed line).

\begin{figure*}
\begin{center}
\includegraphics[height= 9.9cm,width=18.0cm,trim=0.0cm 0.0cm 0.0cm
0.0cm,clip=false]{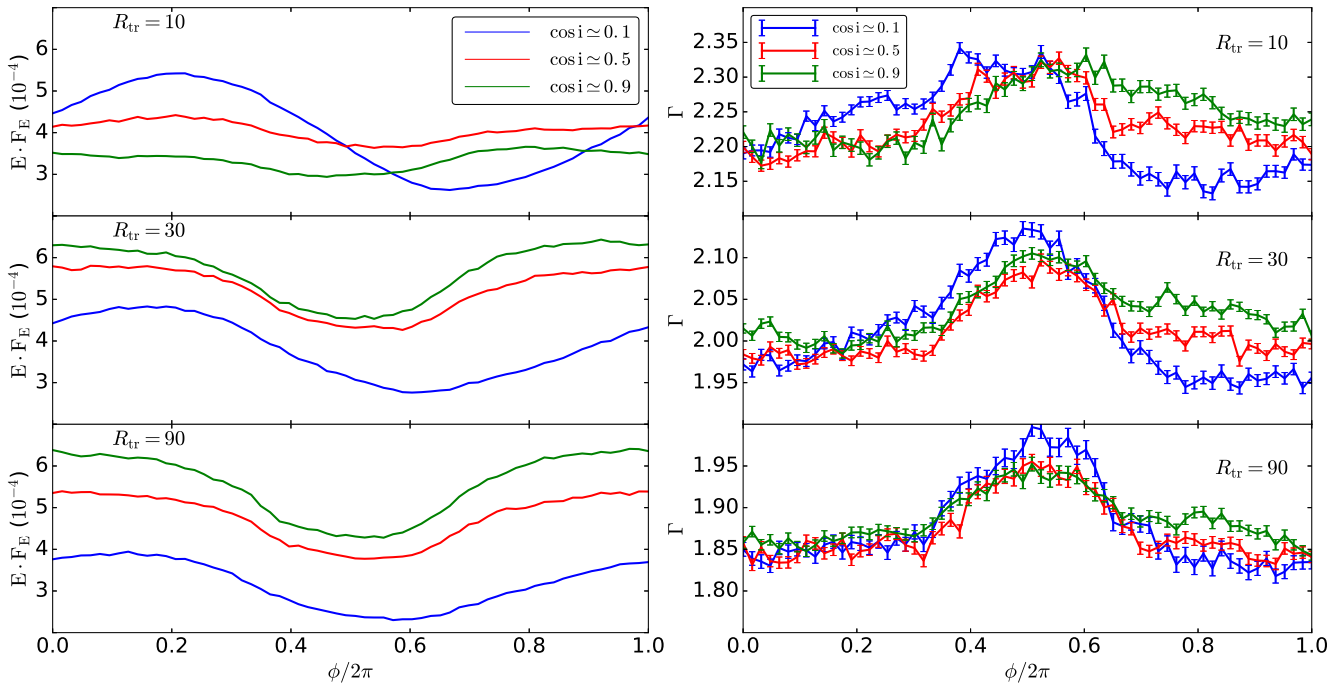}
\end{center}
\caption{
Normalized continuum flux at 4-10 keV (left) and the photon spectral indices (right), as a function of the precession phase for half-opening angle $\theta_{\rm o} = 15^{\circ}$ and the black hole spin $a=0.3$. The top, middle, and bottom panels are for the truncation radius $R_{\rm tr} = 10, 30$, and 90, respectively. In each panel, blue, red, and green solid lines are for the sinusoidal value of viewing angle, $\cos i$ = 0.1, 0.5, and 0.9. On the right, the X-ray photon index $\Gamma$ is derived by the power-law fits to the continuum spectrum, and the error bars in the represent the uncertainties.
\label{slope_0p3_t15}}
\end{figure*}

As mentioned in the previous sections, when the Comptonizing hot flow precesses, the variation of the heating-to-cooling ratio will result in the change of the electron temperature. This effect, combined with the variation of the projected area of the torus due to the precession of the flow, will essentially affect the X-ray spectrum.
In Ingram et al. (2016), the QPO phase-resolving method invented by Ingram et al. (2015) was applied to the combined \textit{X-ray Multi-Mirror Mission} (\textit{XMM-Newton}) and \textit{Nuclear SpecTroscopic ARray} (\textit{NuSTAR}) data from the 2014 outburst of H1743-322. Observationally, both the 4-10 keV continuum normalization and power-law index were found to display the statistically significant ($>3\sigma$) modulation along with the QPO phase for the joint data.   
In Fig. \ref{slope_0p3_t15}, we plot the normalized continuum flux over 4-10 keV (left) and the X-ray photon index $\Gamma$ (right), which is derived by the power-law fits to the continuum spectrum, as a function of the precession phase, for different truncation radii. The half-opening angle $\theta_{\rm o} = 15^{\circ}$. The blue, red, and green solid lines are for $\cos i$ = 0.1, 0.5, and 0.9, respectively.
The waveform-like modulation of the 4-10 keV flux is indeed seen in our simulations, especially when the disk is truncated at small radius ($R_{\rm tr} = 10$) and the system is viewed edge-on. 
Regarding the spectral index, when $R_{\rm tr} = 90$ the spectra are approximately softest at the middle 
phase, when the torus is aligned with the disk (see Fig. \ref{schematic_torus}), while 
the spectra are hardest at phase 0. 
As $R_{\rm tr}$ decreases, e.g., for $R_{\rm tr} = 10$, the overall Comptonizing flow approaches the black hole, so that the continuum shape will be distorted by the relativistic effect, leading to the distortion of the evolution of the photon index. 

\begin{figure*}
\includegraphics[width=2\columnwidth]{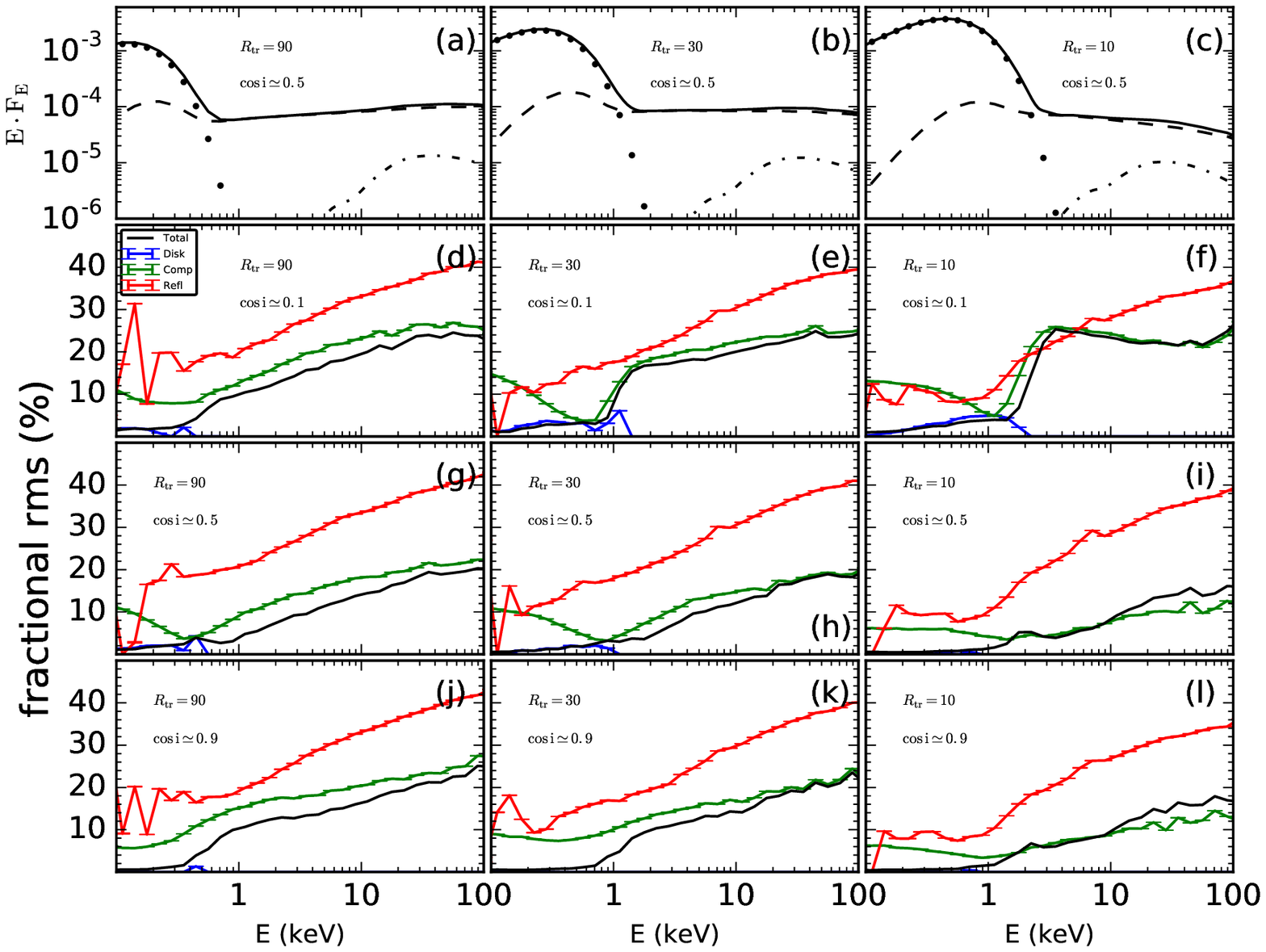}
\caption{
Normalized radiation spectra and the fractional $\rm rms$ spectra for the half-opening angle of the flow $\theta_{\rm o} = 15^{\circ}$ and black hole spin $a=0.3$. The left, middle, and right panels are for the truncation radius $R_{\rm tr}$ = 90, 30, and 10, respectively. The top panels are the phase-averaged energy spectra, which consist of disk (dotted line), Comptonization (dashed) and reflection (dashed-dotted) emission components. The second, third, and fourth rows are the fractional rms spectra that correspond to the sinusoidal value of viewing angle $\cos i$ = 0.1, 0.5, and 0.9, respectively. In each subplot of the fractional rms spectrum, the blue, green, and red lines represent the fractional rms for disk, Comptonization, and reflection alone, respectively, while the black solid line represents the variability of the total emission.
\label{rms_0p3_t15}}
\end{figure*}

In order to have a close look at the X-ray variability in this model, we calculate the energy-dependent fractional rms based on the radiation spectra at different phases over one complete precession of the flow. The results are plotted in Fig. \ref{rms_0p3_t15}, where the left, middle, and right panels are for $R_{\rm tr} = 90, 30,$ and 10, respectively. The top panels show the phase-averaged spectra and their decomposition into the disk, Comptonization, and reflection components, including the iron emission line. 
In this paper, the disk is assumed to be neutral with zero ionization for simplicity. Note that only a small fraction of photons escaping from the central source can illuminate the disk, so that the amplitude of the reflected component is low.
Consequently, the iron line is most invisible in the overall energy spectrum. 
The second, third, and fourth rows are the fractional rms spectra as a function of energy for $\cos i$ = 0.1, 0.5, and 0.9, respectively. 
Throughout this paper, the uncertainties on the fractional rms are estimated based on Equation (B2) of Vaughan et al.(2003). The results are as follows: (i) The observed disk radiation (blue lines) is not variable in most cases, except when the source is in the soft state being observed with a large viewing angle (panel (f)). As for the latter case, the fractional rms variability amplitude of the disk component can be up to about 5\%, which is mainly due to the occultation of the photons emitted from the disk behind the torus. 
As the viewing angle decreases (viewing closer to face-on), the overlapping region and the resultant variability are reduced. The disk emission will gradually be more variable as the truncation radius moves inward, corresponding to the spectral evolution from hard to soft state. (ii) At $E$ > 3 keV, where the Comptonized emission dominates over other components, the fractional rms has a flat distribution as a function of energy (e.g., $R_{\rm tr} = 10$, in the soft state and being viewed edge-on), or increases with the energy (in the hard state with $R_{\rm tr} = 90$). 
In the intermediate/soft state ($R_{\rm tr} = 10, 30$), the fractional rms spectra of the Comptonized emission are concave at the low energy $E < 3$ keV, which
corresponds to the unscattered photons from the disk traveling through the torus, at the low energy $E < 3$ keV. Therefore, the pivot energy is associated with disk peak temperature. 
(iii) The reflection component is highly variable, with the fractional rms amplitude reaching up to 30\%. The fractional rms spectra, especially at high photon energy, are independent of the viewing inclination, in both hard and soft state.
(iv) At the low energy where the total spectrum is dominated by the disk radiation, the fractional rms is very low, $< 1\%$, while the variability amplitude of the Comptonized emission from the torus is high, $\sim 10\%$; on the other hand, where the total spectrum is dominated by the Comptonized radiation rather than the disk emission, the fractional rms spectrum of the total radiation has the most contribution from the former, i.e., flattening or increasing with photon energy.

\begin{figure*}
\includegraphics[width=2\columnwidth]{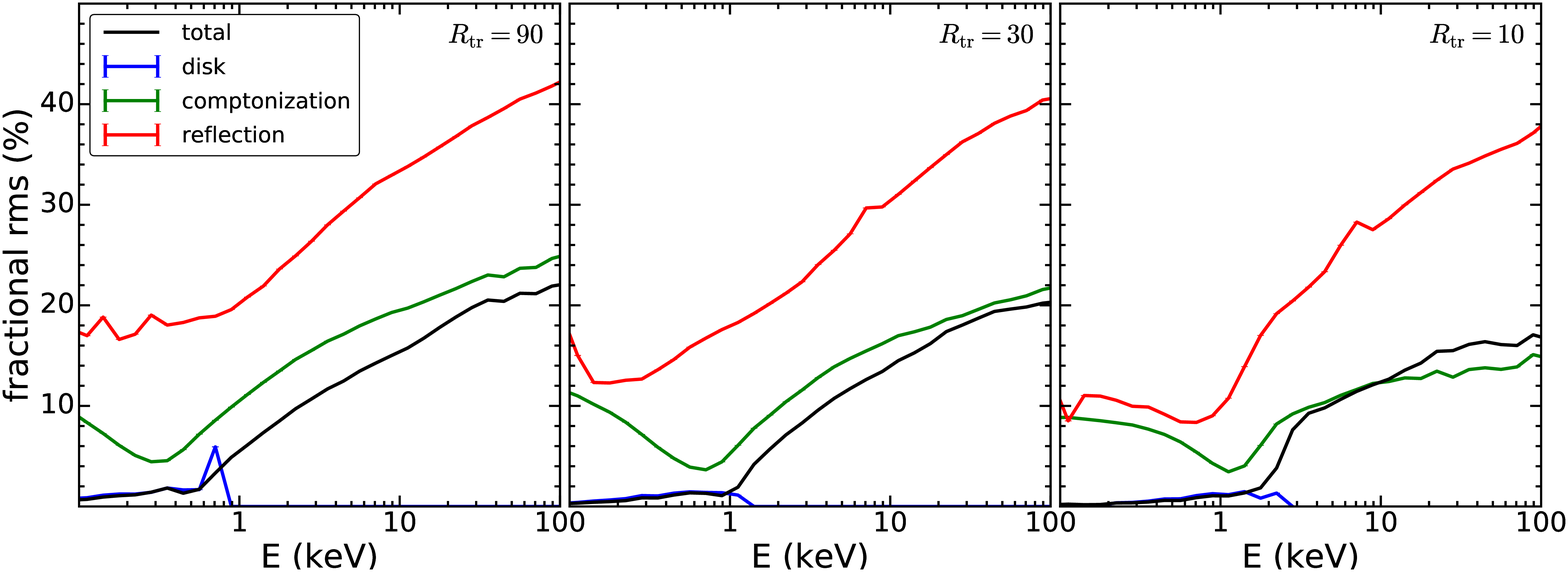}
\caption{
Fractional $\rm rms$ spectra of the radiation integrated over all inclination angles for the half-opening angle of the flow $\theta_{\rm o} = 15^{\circ}$ and the black hole spin $a=0.3$. The left, middle, and right panels are for truncation radius $R_{\rm tr}$ = 90, 30, and 10, respectively. 
In each subplot of the fractional rms spectrum, the blue, green, and red lines represent the fractional rms for disk, Comptonization, and reflection alone, respectively, while the black solid line represents the variability of the total emission.
\label{rms_0p3_t15_incli}}
\end{figure*}

In Fig. \ref{rms_0p3_t15_incli}, we plot the fractional $\rm rms$ spectra of the radiation integrated over all inclination angles, for different truncation radii. This enables us to get rid of the inclination effects dependent on the observer position as mentioned above, hence isolating the spectral variability caused by the changing geometry of the system. The variability amplitudes and the fractional rms spectrum for both the Comptonization and the reflection components evolve during the spectral transition as the truncation radius decreases from $R_{\rm tr}$ = 90 to $R_{\rm tr}$ = 10.

\subsubsection{Effects of varying Black Hole spin}

In this case, we study the spectral and variability properties for black hole spin $a = 0.3, 0.6,$ and 0.9, while the truncation radius is fixed at $R_{\rm tr} = 10$. 

\begin{figure*}
\includegraphics[width=2\columnwidth]{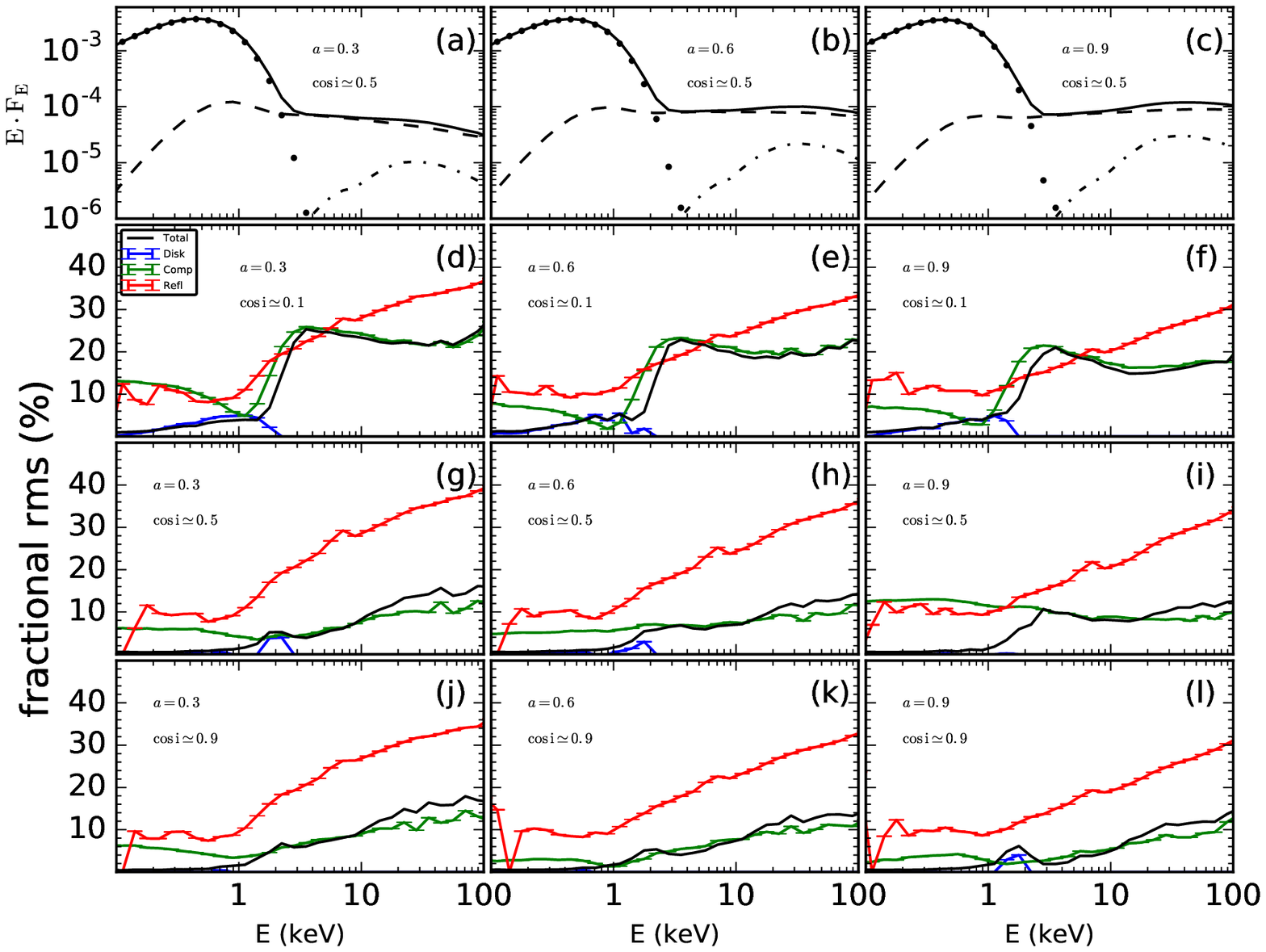}
\caption{
Normalized radiation spectra and fractional $\rm rms$ spectra for the half-opening angle of the flow $\theta_{\rm o} = 15^{\circ}$ and the truncation radius $R_{\rm tr} = 10$. The left, middle, and right panels are for the black hole spin $a = 0.3, 0.6,$ and 0.9, respectively. The top panels are the phase-averaged energy spectra that consist of disk (dotted line), Comptonization (dashed), and reflection (dot-dashed) emission components. The second, third, and fourth rows are fractional ${\rm rms}$ that corresponds to the sinusoidal value of viewing angle $\cos i$ = 0.1, 0.5, and 0.9, respectively. In each subplot of the fractional rms spectrum, the blue, green, and red lines represent the fractional rms for disk, Comptonization, and reflection alone, respectively, while the black solid line represents the variability of the total emission.
\label{rms_10rg_t15}}
\end{figure*}

The fractional rms spectra 
are plotted in Fig. \ref{rms_10rg_t15}, where the
left, middle, and right columns are for $a = 0.3, 0.6,$ and 0.9, respectively. The top panels show the phase-averaged spectra and their decomposition into the disk, Comptonization, and reflection components. The second, third, and fourth rows are the fractional rms spectra as a function of energy for $\cos i = 0.1, 0.5,$ and 0.9, respectively. 
Comparing the fractional rms spectra for different black hole spin (from the left to right column), 
it can be seen that the variability of the disk emission does not depend on the spin, while variability amplitude of the Comptonized emission is counterintuitively suppressed as the spin increases. Correspondingly, as the output of the X-ray reprocessing in the disk, the variability of the reflection component is also suppressed for large black hole spin.

\subsection{Thick-torus case}
The half-opening angle of the torus is uncertain, and physic consideration and simulation are beyond the scope of our current work. Therefore, we consider another possibilities by simply assuming the value of the half-opening angle. 
In this section, we repeat previous analysis, but in the case of the half-opening angle $\theta_{\rm o} = 45^{\circ}$. Other parameters are fixed with the same values as in the case of $\theta_{\rm o} = 15^{\circ}$.

\subsubsection{Effects of varying the truncation radius}

\begin{figure*}
\includegraphics[width=2\columnwidth]{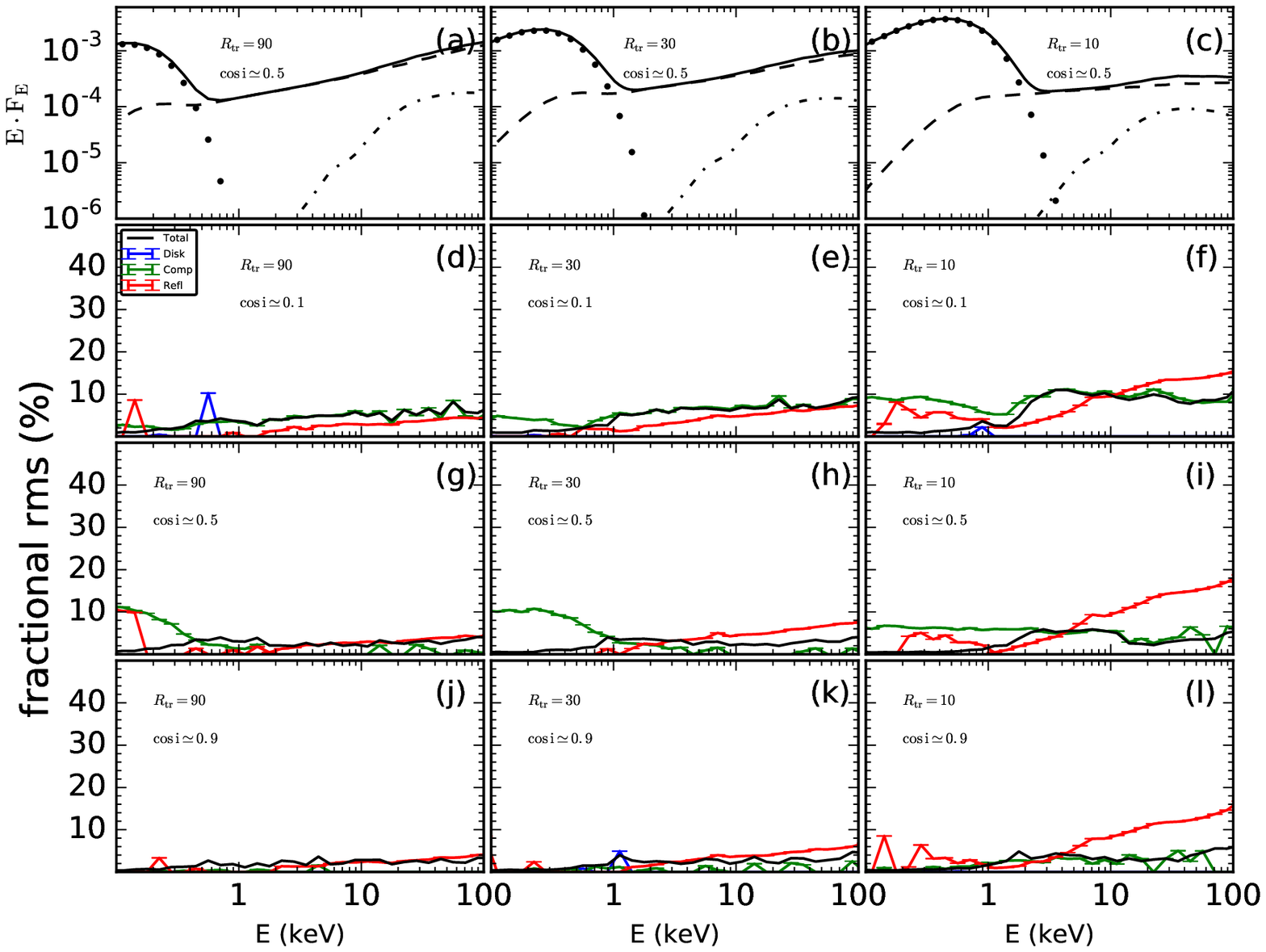}
\caption{
Normalized radiation spectra and the fractional $\rm rms$ spectra for the half-opening angle of the flow $\theta_{\rm o} = 45^{\circ}$ and black hole spin $a=0.3$. The left, middle, and right panels are for the truncation radius $R_{\rm tr} = 90, 30$, and 10, respectively. The top panels are the phase-averaged energy spectra that consist of disk (dotted line), Comptonization (dashed), and reflection (dot-dashed) emission components. The second, third, and fourth rows are fractional ${\rm rms}$ that corresponds to the sinusoidal value of viewing angle $\cos i$ = 0.1, 0.5, and 0.9, respectively. In each subplot of the fractional rms spectrum, the blue, green, and red lines represent the fractional rms for disk, Comptonization, and reflection alone, respectively, while the black solid line represents the variability of the total emission.
\label{rms_0p3_t45}}
\end{figure*}

The fractional rms spectra are plotted in Fig. \ref{rms_0p3_t45}, where  the
left, middle, and right columns are for $R_{\rm tr} = 90, 30,$ and 10, respectively. The top panels show the phase-averaged spectra and their decomposition into the disk, Comptonization, and reflection components. The second, third, and fourth rows are the fractional rms as a function of energy for $\cos~i$ = 0.1, 0.5, and 0.9, respectively. It can be seen that (i) comparing with the results in Fig. \ref{rms_0p3_t15}, the variability amplitudes for $\theta_{\rm o} = 45^{\circ}$ are overall smaller than the ones for the small opening angle; (ii) the disk emission is invariable regardless of the spectral states; (iii) as the source evolves from hard state to soft state with the truncation radius decreasing from 90 to 10, the variability amplitudes of the Comptonized emission from the torus are enlarged;
(iv) the variabilities of the reflection radiation are enhanced, with the fractional rms reaching up to 20\%; and (v) as the viewing angle decreases from edge-on to face-on, the overall variabilities of the source are suppressed.


\subsubsection{Effects of varying the BH spin }



\begin{figure*}
\includegraphics[width=2\columnwidth]{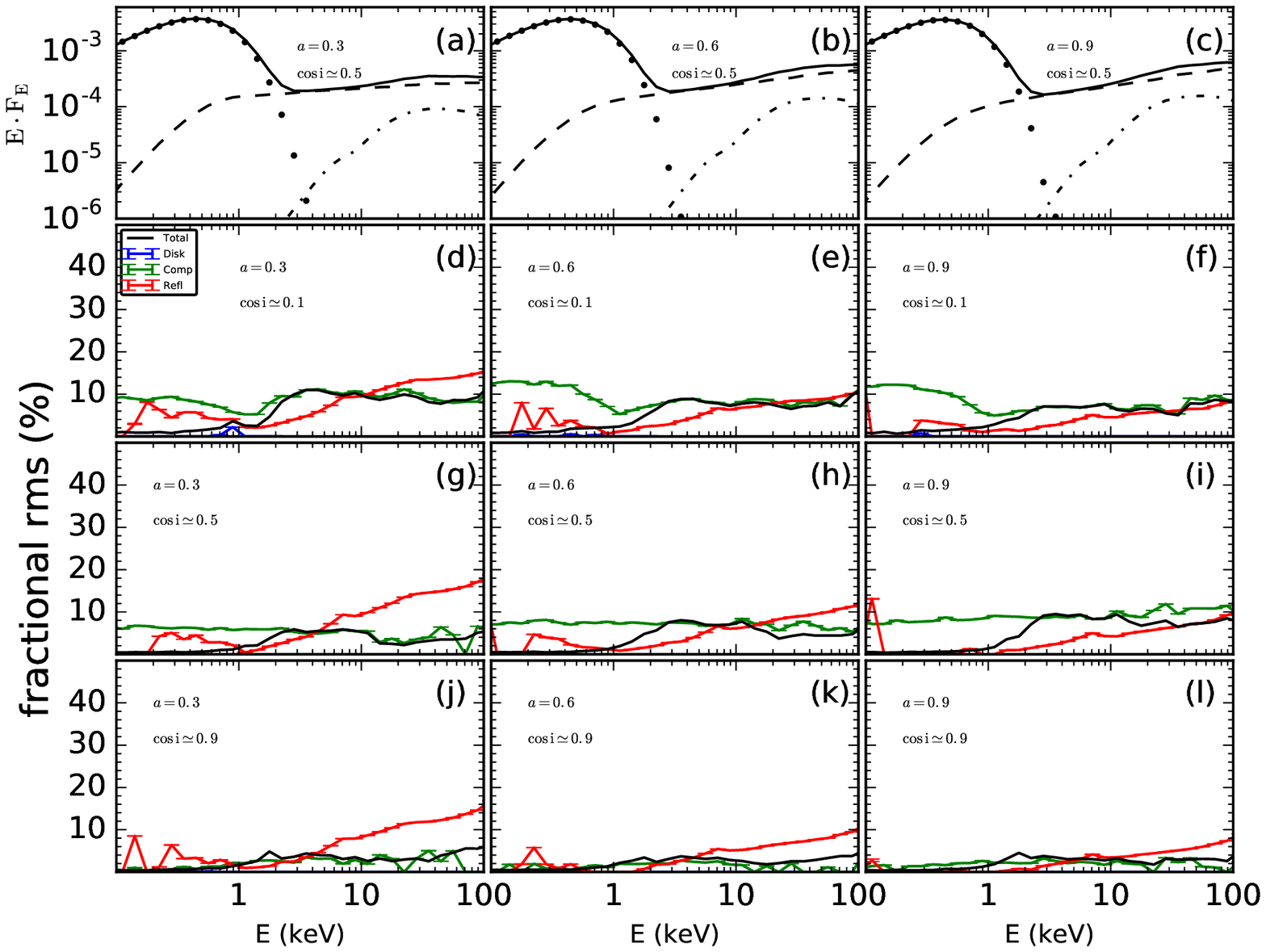}
\caption{
Normalized radiation spectra and fractional $\rm rms$ spectra for the half-opening angle of the flow $\theta_{\rm o} = 45^{\circ}$ and the truncation radius $R_{\rm tr} = 10$. The left, middle, and right panels are for the black hole spin $a = 0.3, 0.6,$ and 0.9, respectively. The top panels are the phase-averaged energy spectra that consist of disk (dotted line), Comptonization (dashed), and reflection (dot-dashed) emission components. The second, third, and fourth rows are fractional ${\rm rms}$ that corresponds to the sinusoidal value of viewing angle $\cos i$ = 0.1, 0.5, and 0.9, respectively. In each subplot of the fractional rms spectrum, the blue, green, and red lines represent the fractional rms($E$) for disk, Comptonization, and reflection alone, respectively, while the black solid line represents the variability of the total emission.
\label{rms_10rg_t45}}
\end{figure*}

The energy-dependent fractional rms($E$) based on the spectra at different phases within the complete period of the flow precession are plotted in Fig. \ref{rms_10rg_t45}, where  the left, middle, and
right columns are for $a = 0.3, 0.6,$ and 0.9, respectively. The top panels show the phase-averaged spectra and their decomposition into the disk, Comptonized, and reflection components. The second, third, and fourth rows are the fractional rms($E$) as a function of energy for $\cos~i$ = 0.1, 0.5, and 0.9, respectively. The fractional rms($E$) of the Comptonized radiation is constant across high photon energy, and the amplitude is independent on the black hole spin, while the fractional rms spectra of the reflection components increase with photon energy. However, the variability amplitude of the latter decreases as the spin increases. Although the fractional rms spectra of the reflection components are independent of the viewing angle, the fractional rms spectra of the Comptonized radiation are larger when the source is viewed with large inclination. Therefore, the fractional rms spectra of the total radiation increase with the viewing angle. 


\section{Discussion}

In this work, we develop a relativistic Monte Carlo code, to compute the Compton-scattered X-ray flux produced by the hot inner flow intercepting seed photons from an outer truncated standard disk. The inner flow will undergo the Lense-Thirring precession if the rotation axis of the flow misaligns with the black hole spin axis owing to the frame-dragging effect. It was proposed that the precession of the inner flow gives rise to the modulation of the X-ray flux through the combination of the geometrical effect, e.g., the self-occultation, the wobbling of the projected area, and the relativistic effect (Ingram et al. 2009).
We investigate the energy-dependent X-ray variabilities from the Lense-Thirring precession in the truncated disk geometry and, furthermore, the evolution of the variability properties during the spectral transition from the hard to soft state, which is implemented by the decrease of the truncation radius of the outer disk toward ISCO. 


\subsection{Mechanism of Low-frequency QPO variability}
Before discussing the influence of model parameters on the QPO variability for disk, Comptonization, and reflection emission, we should explain the fundamental mechanism of QPO variability for these spectral components. The evolution of the QPO variability along with spectral transition will be discussed in the next subsection.

{\bf \it{Disk thermal emission variability}:}
The modulation of disk emission is seen in our simulation for some particular cases, e.g., the blue line in panel (f) of Fig. \ref{rms_0p3_t15}, where the source is in the soft state and observed with large inclination angle. In this case, the disk truncation radius
is small, with $R_{\rm tr} = 10$ close to the torus inner radius $R_{\rm in} = 5$ for $a=0.3$, and thus the extent of the torus is small. This will result in parts of the innermost disk behind the torus being observed or hidden from view during the precession phase, which depends on the torus position. This changing appearance of the occulted disk then leads to the inner disk emission being variable (i.e., the bump at energy $E$ < 3.0 keV), due to not only the change in the occulted area of the disk but also the intrinsic blueshift/redshift modulation of the rotating disk emission. 

{\bf \it{Comptonized emission variability}:}
The variability of the Comptonization emission is intrinsic in the Lense-Thirring precession mechanism. The precession flow emitting the X-ray is translucent with the optical depth in the range of $\sim 1-2$, 
which is determined by the \textit{eqpair} code
, so that the varying of the projected area of the flow due to precession will contribute to the flux modulation. In other words, the angle between the line of sight and the axis of the torus changes with precession
phase, so that the optical depth along the line of sight varies as the torus wobbles, which leads to the variable fraction of scattered and unscattered photons and thus the modulation of X-ray emission. Furthermore, photons with higher energy undergo a higher order of Comptonization before escaping the torus, so that the effect of the wobbling of the torus on the X-ray modulation is more significant for higher-energy photons. This then explains the increase of the fractional rms with photon energy (e.g., green lines in left panels of Fig. \ref{rms_0p3_t15}). In addition, the electron
temperature changes, if the relative geometry of the torus and outer disk changes during the precession phase, because of the change of heating-to-cooling ratio of the plasma. This leads to spectral (slope) variations and contributes to the extra variability. 

{\bf \it{Reflection component variability}:}
In our simulation, the phased-averaged reflection components including the iron line are weak with respect to the Comptonization component (e.g., Fig \ref{rms_0p3_t15}). This is because only a small fraction of photons escaping from the central source can illuminate the disk, so that the amplitude of the reflected component is low. 
Furthermore, it has been shown in the literature that the ionization parameter plays an important role in shaping the resultant iron emission line (e.g, Garc{\'{\i}}a et al.(2013), Basak \& Zdziarski(2016). In this work, for simplicity, the disk is assumed to be neutral with zero ionization. Therefore, a more realistic ionization parameter would return an iron line more consistent with observations.

Although the phased-averaged reflection components are weak, the fractional rms amplitudes of the reflection components increase with photon energy in most of the simulation results. As indicated in Ingram
\& Done (2012), the tilted flow illuminates different azimuths
of the disk as it precesses. The resulting reflection emission
is blueshifted when the flow predominantly irradiates the approaching region
of the rotating disk and redshifted when the flow predominantly irradiates
the receding region of the disk.
Therefore, this anisotropical irradiation due to the nonspherically symmetric geometry of the precessing torus
intrinsically accounts for the reflection variability considered here. 
Because of the torus configuration,
the closer to the innermost region of the torus the irradiation source is,
the more anisotropic the irradiation with respect to the disk is.
Moreover, compared to the irradiating photons escaping from the outer region of the torus,
the energy of the ones from the innermost region is high, so that, the fractional rms amplitude of the reflection components 
due to the precessing illumination increases with photon energy.
Moreover, there are interesting features in the soft states, i.e., a spike at $E$ $\sim$ 6-8 keV (e.g., right panels in Fig. \ref{rms_0p3_t15}). This arises from the variation of the Fe k$\alpha$ line, which is the characteristic signature for the reflection components. We find that both the  centroid energy and line width of the iron line vary as the flow precesses, which agrees with the results of Ingram \& Done (2012). The results will be presented in detail in the next paper.
Finally, the fractional rms($E$) for the total radiation is determined depending on the contribution of the flux from individual spectral components. For example, in  panel (d) of Fig. \ref{rms_0p3_t15}, at soft energy $E$ < 1keV where the disk emission dominates over other spectral components, the fractional rms amplitude of the total radiation is very small, as the disk emission is weakly variable, regardless of the high variability of the Comptonization and reflection components.

\subsection{Evolution of QPO variability}

Based on the mechanism of the QPO variability, in this section we will investigate the results of Fig. \ref{rms_0p3_t15} to study the evolution behavior of the QPO variability along with the spectral transition. The discussion is divided into two subcases for small and large viewing angles.

\subsubsection{Small viewing angle}
When the viewing angle with respect to disk axis is small, e.g., the source being viewed face-on (bottom panels), the patterns of the fractional rms spectra are simple. The fractional rms spectra for the Comptonization, reflection, and hence total components all monotonously increase with energy at $E$ > 1 keV, e.g. panel (j). 

The ratio of the disk luminosity intercepted by the torus to the total disk luminosity and the resulting electron temperature are plotted in Fig. \ref{lsoft}. In the hard state, e.g., $R_{\rm tr} = 90$ or 30, the electron temperature varies ($\sim 10\%$) during the precession phase, while in the soft state, e.g., $R_{\rm tr} = 10$, there is relatively weak change ($\sim 6\%$) in the temperature during the precession. Therefore, this is why the variability amplitudes of the X-ray emission (both the Comptonization and reflection components) in the soft state are relatively lower than the ones in the hard state, comparing panel (j) with panel (l).


\subsubsection{Large viewing angle}
As explained in Sect. 4.1, when the viewing angle with respect to the disk axis is large, e.g., being viewed edge-on, 
the occultation of the disk by the torus becomes important.
In the hard state, the disk is truncated at large radius, and thus the inner hot flow is spatially extensive, so that the portion of the disk that is occulted by the inner torus is large and not variable during the precession. Therefore, almost constant disk emission around peak energy $E\sim$0.1 keV would be expected, e.g., panel (d). 
Besides of the varying of the optical depth along the line of sight and plasma temperature, which contributes to the variabilities of the Comptonization, the self-occultation effect should also give rise to the flux modulation for edge-on observation. That is to say, when the torus terminates at large radius, the inner part will be totally obscured by the outer boundary at particular precession phase, while the the outer boundary can be always be observed during the precession. 

As the source evolves into soft state (panel (e) and (f)) with the
inner radius of the disk moving inward, the pattern of the fractional 
rms spectra is complex: (1) below 3 keV, the bump
spectrum for the disk component (blue line) and the concave spectrum
for the Comptonization component (green line); (2) at
high energy $E > 3$ keV, roughly flat distribution for the Comptonization component. As the truncation radius decreases, the
smaller and smaller extent of the torus and the light-bending effect due to strong gravity will result in parts of the innermost disk behind the torus being observed or hidden from
view during the precession phase, which depends on the torus
position. This changing appearance of the unocculted disk
then leads to the inner disk emission being variable, namely,
the bump of the fractional rms spectrum for the disk component at $E < 3$ keV, due to not only the change in
the occulted/unocculted area of the disk but also the intrinsic blueshift/redshift of the rotating disk emission. The concave fractional rms spectra (green line) correspond to the
bump of the energy spectrum of the Comptonization component at low energy (panel
(b) and (c)) which is attributed to those photons from the occulted
disk traveling through the torus but without Comptonization
(note that in our simulation, such unscattered photons are
saved as the emission from the torus). Therefore, the changing
appearance of the occulted disk also naturally contributes to
the variability of the torus emission at $E < 3$ keV (unscattered photons). Moreover, during the precession phase, the
outer torus that emits low-energy unscattered photons wobbles more significantly than the inner torus that emits relatively high energy unscattered photons, so that the resultant
fractional rms due to the wobble decreases with photon energy, namely,
the left wing of the concave fractional rms spectrum for the Comptonization component. As for the fractional rms
spectra at high energy $E > 3$ keV that originate from the
scattered photons, its amplitude is about 30\%. The soft state in the truncated disk model requires that the torus is spatially compact, meaning that low-
and high-energy photons are emitted in the close region of the
torus. Therefore, the fractional rms of the Comptonized
photons with $E > 3$ keV should be somewhat independent of
photon energy. Finally, for the fractional rms spectrum in the range of
1.0 and 3.0 keV, there is an increasing contribution from the
scattered photon, so that the fractional rms will increase with
photon energy, which is shown as the right wing of the concave fractional rms spectra of the Comptonization component.


\subsection{Dependence of the QPO variability}

In this section we will discuss the role of the major parameters in shaping both the QPO variability and its evolutional behavior along the spectral transition. The corresponding results are plotted in Fig. \ref{rms_10rg_t15}-\ref{rms_10rg_t45}.  

\subsubsection{On black hole spin}

Regarding the potential influence of the black hole spin on the spectral and timing properties, it might be implemented by affecting the accretion physics in the innermost region near the rotating black hole and modifying the emergent spectrum due to gravitational redshift and the light-bending effect. Studying the role of the strong gravity of a rotating black hole on the accretion flow is beyond the scope of this work. It is suggested by analytic and simulation works that the misalignment of the accretion flow will result in not only the precession of the entire flow as a solid body but also the additional torque from a black hole, contributing to the stresses (Bardeen \& Petterson 1975; Nelson \& Papaloizou 2000; Fragile \& Anninos 2005). Then, the increased stresses will increase inward velocity and therefore decrease the density of the flow (Fragile et al. 2007; Liska et al. 2018). In this sense, the inner radius of the precessing flow will be larger than the ISCO, which is usually assumed as the inner radius of the disk.
In Fig. \ref{rms_10rg_t15}, the evolution of both phase-averaged energy spectra and energy-dependent fractional rms for different black hole spins is shown, where the truncation radius of the outer disk is fixed at $R_{\rm tr} = 10$. The inner radius of the inner flow is simply assumed in the same way as in Ingram et al. (2009). The relativistic correction to the emergent spectrum is internally implemented in our simulation by tracing each individual photon from the flow to the infinity.

When the viewing angle with respect to disk axis is small, e.g., viewing face-on (bottom panels), there is no self-occultation effect, so that the combination of the change in optical depth  along the line of sight due to the wobbling of the torus and the relativistic effects leads to the monotonous increase in the fractional rms spectra of both the Comptonization and the resulting reflection components, as a function of photon energy. The resultant variation in plasma temperature (due to the chang in the number of the seed soft photons interacting with the torus) shows some subtle difference among different spins, e.g., the red lines in Fig. \ref{lsoft}. Therefore, no obvious change in the fractional rms amplitudes of the Comptonization components would be expected, e.g., green lines in panel (j)-(l) of Fig. \ref{rms_10rg_t15}. Yet, it is found that the fractional rms amplitudes of the reflection component increase as the spin decreases. As explained in the previous section, the reflection variability is attributed to the blueshift/redshift modulation in response to the precession of the illuminating flow. As the spin decreases, the inner torus extends further inward, so that part of the illuminating photons will arise from the region closer to the black hole. As the flow precesses, the illuminating spectrum is highly modulated owing to the relativistic effect of the strong gravitational field, contributing to extra variability amplitude of the resultant reflection radiation. This is why the reflection variability for the low spin is larger than the one for high spin.

When the viewing angle with respect to the disk axis increases, e.g., from panel (j) to panel (d) in Fig. \ref{rms_10rg_t15}, the self-occultation effect plays an increasing role in shaping the fractional rms spectra. Especially when the source is viewed edge-on (panel (d)), the fractional rms flattens across high energy. In this case of large viewing angle, as the spin increases, from panel (d) to panel (f), the torus recedes, and hence the extent of the torus decreases, so that the self-occultation effect is reduced partly as a result of the light-bending effect. Therefore, the variability amplitude of both the Comptonization and reflection components is suppressed in the case of high spin.

\subsubsection{On the torus opening angle}

In our simulation, model parameters regarding the configuration of the geometry, e.g., the misalignment angle of the black hole spin with respect to the the symmetry axis of the disk $\theta_{\rm p}$, the misalignment angle of the symmetry axis of the torus with respect to the spin $\theta_{\rm t}$ and the half-opening angle of the torus, currently cannot be constrained well by the observational and theoretical studies. 
Note that $\theta_{\rm o}$ < $\theta_{\rm p}$ would probably give very different results, since the disk would see the `bottom' of the inner flow at some precession phases.
It is impossible to explore the effect of all model parameters on the spectral and timing properties for the transient source in this paper. 
Therefore, in this paper, we investigate the spectral and variability properties in the Lense-Thirring precession model for two exemplary cases of the opening angle, i.e., $\theta_{\rm o} = 15^{\circ}$ and $45^{\rm \circ}$.

In Fig. \ref{rms_0p3_t45}, the results for different truncation radii in the case of the half-opening angle $\theta_{\rm o} = 45^{\circ}$ are shown, in the comparison to the results in Fig. \ref{rms_0p3_t15}. The energy spectrum evolution during the state transition is similar to that in the case of $\theta_{\rm o} = 15^{\circ}$, while the fractional rms spectra are obviously different. 
The overall variability amplitudes of the Comptonization component are highly suppressed, $\leq 10\%$, irrespective of the spectral state. 
As the torus becomes thick with large opening angle, the torus is geometrically thicker than the range of its wobble due to precession, so that the variation of the projected area of the wobbling torus will be reduced. Moreover, the variation of the plasma temperature during the precession for the thick torus is subtle owing to an almost constant heating-to-cooling ratio of the plasma. The overall effect will result in the weak variability of the Comptonization radiation.

Yet, it is found that the variability of the reflection component is enlarged from the hard to soft state, e.g, comparing left panels with right panels in Fig. \ref{rms_0p3_t45}. In the hard state with large truncation radius, the torus is spatially extensive, so that the anisotropy of the irradiation by the precessing torus with respect to the approaching/receding region of the disk is subtle, which leads to the weak variability.
However, when the source evolves toward the soft state, the extent of the precessing torus decreases. Then, the effect of the anisotropical irradiation by the precessing torus becomes significant. This triggers the blueshift/redshift modulation for the reflection spectrum, namely, enhancing the reflection variability.

In Fig. \ref{rms_10rg_t45}, the results of the spectral and variability properties are shown for different spins in the case of the half-opening angle $\theta_{\rm o} = 45^{\circ}$. Again, the overall amplitudes of the fractional rms spectra of both the Comptonization and total radiation are roughly constant and significantly lower than the ones for $\theta_{\rm o} = 15^{\circ}$, which has been discussed above. Moreover, the fractional rms spectra of the Comptonization are irrespective of the spin, because of the combination of large torus thickness and a subtle change in the electron temperature.
And comparing to the case of small half-opening angle, it is more obvious that the fractional rms amplitudes of the reflection component increase as the spin decreases.

\subsection{Comparison with observations}

The aim of this work is to single out the effects of X-ray modulation due to Lense-Thirring precession of the hot flow on the X-ray variability spectra. Given that we are discarding other mechanisms involving instabilities in the flow, at present our simulations allow only for a qualitative comparison with observations.

\begin{figure}
\includegraphics[height=13.0cm,width=21.0cm,trim=0.0cm 0.0cm 0.0cm
 0.0cm,clip=true]{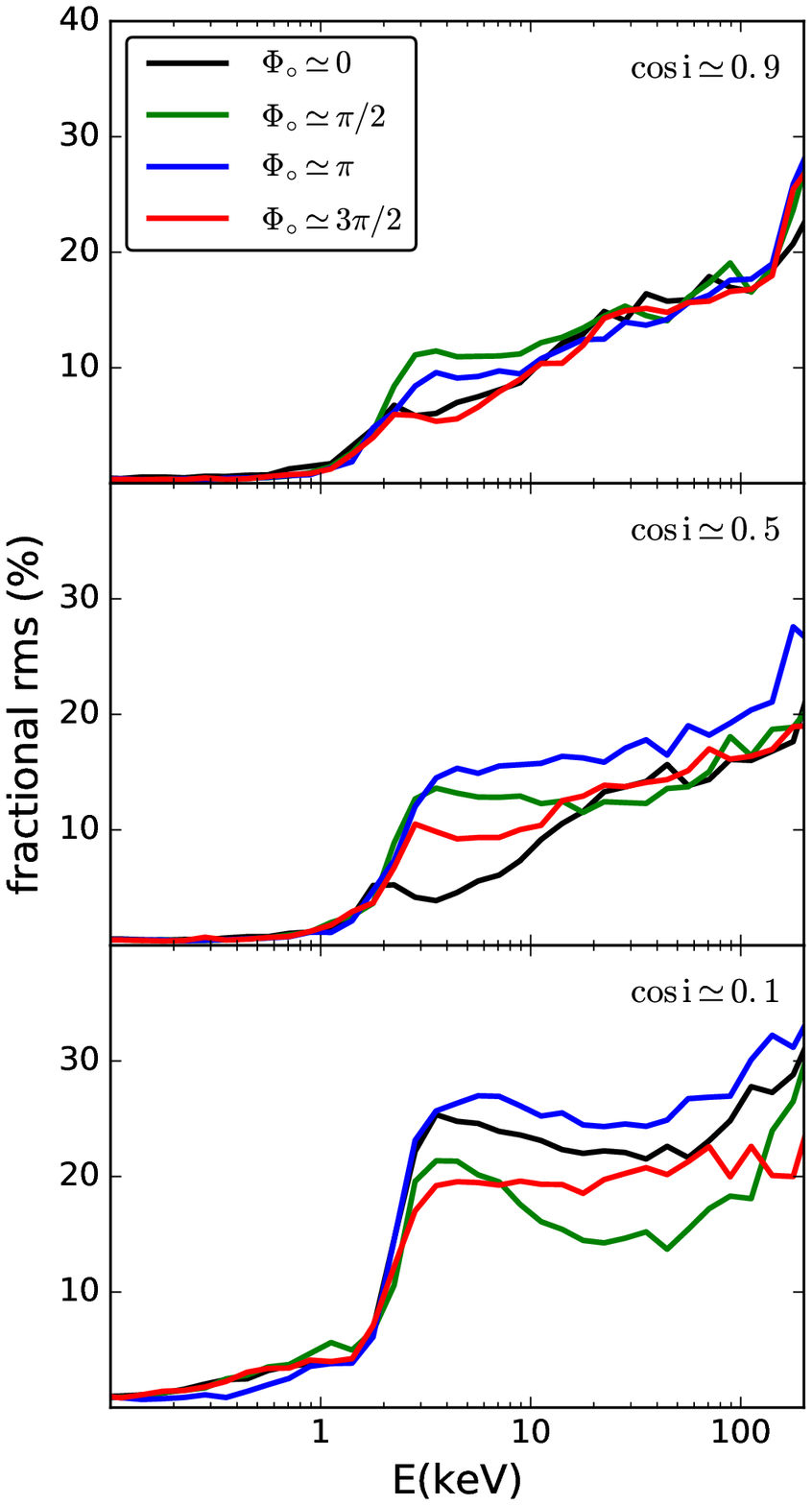}
\caption{
Fractional $\rm rms$ spectra for the half-opening angle of the flow $\theta_{\rm o} = 15^{\circ}$. The truncation radius $R_{\rm tr}$ = 10, and the black hole spin $a=0.3$. The top, middle, and bottom panels correspond to the sinusoidal value of the viewing angle $\cos i$ = 0.9, 0.5, and 0.1, respectively. In each panel, the black, green, blue, and red lines are for the viewer azimuth $\Phi_{\circ} = 0, \pi/2, \pi$, and $3\pi/2$, respectively.
\label{rms_0p3_t15_phi}}
\end{figure}

The fractional rms spectra as a function of energy could provide information and constraint on the variability mechanism and the accretion physics in the X-ray binaries.
Gierli{\'n}ski \& Zdziarski (2005) analyzed several observations of two X-ray binaries in various states from the Rossi X-ray
Timing Explorer ($RXTE$), and then extracted fractional rms spectra from the PCA. The presented characteristic patterns of the fractional rms clearly show the evolution during the state transition. However, those fractional rms spectra were
integrated over the wide range of Fourier frequency, while
the fractional rms spectra that are simulated in our paper correspond to the QPOs in the $<$ 10 Hz region. The work of investigating the fractional rms spectra for the QPOs within $1-10$ Hz was done by Sobolewska \& {\.Z}ycki (2006), which showed that the fractional rms decreases with energy for some objects and increase for others.
Note that all the simulated results in Fig. \ref{spec_0p3_t15}-\ref{rms_10rg_t45} including the fractional rms spectra, are for the viewer azimuth $\Phi_{\circ} = 0$. However, since the geometry in this work is inherently asymmetric, it would be expected that the viewer azimuth has a quite profound effect on computing the variability properties. Indeed, Fig. 9 and 10 of Ingram et al (2015) show that the viewer azimuth can have a big influence on the observed fractional rms. In Fig. \ref{rms_0p3_t15_phi}, the fractional rms spectra for different viewer azimuths are plotted, where the truncation radius $R_{\rm tr}=10$ and the spin $a=0.3$. The top,
middle, and bottom panels correspond to low, moderate, and
high inclination, i.e., $cos i \simeq$ 0.9, 0.5, and 0.1, respectively. It can be seen that indeed the fractional rms spectrum varies with the viewer azimuth, even when other fundamental parameters are fixed. Moreover, such an effect is more profound for high inclinations than low inclinations. This means that the comparison of these predictions to observations of individual sources should take the effect of the viewer azimuth into account. 
As shown on the left of Fig. \ref{rms_0p3_t15_phi}, if the object is observed at different (polar) inclination angles and also different viewer azimuths, the fractional rms will display the diversified (energy dependence) variability properties.
Indeed, Fig. 4 of Sobolewska \& Zycki (2006) shows a hard fractional rms spectrum for 4U 1630 and a soft one for XTE J1550. Fig. 6 of Ingram \& van der Klis (2015) shows that GRS 1915+105 has a hard rms spectrum for observations with a hard time-averaged spectrum.

Considering the results in Fig. \ref{rms_0p3_t15} and \ref{rms_0p3_t45} from our simulation, the large half-opening angle gives almost constant fractional rms irrespective of photon energy, while the small one predicts the increasing of the fractional rms with photon energy. Therefore, it seems that the half-opening angle of the precessing torus decreases as the source evolves toward to the soft state, so that the rms gradually increases with energy. This postulated behavior of the half-opening angle during the state transition somewhat is somewhat plausible, provided that the height of the torus (inner hot flow) is supported by the magnetic pressure and/or radiation pressure, which might correlate with the plasma temperature; meanwhile, it was found that the high-energy cutoff in the X-ray spectrum of GX 339-4 decreases during the spectral transition from hard to soft state (Belloni et al. 2011).


Recently, Motta et al. (2015) performed a systematic study of fast time-variability properties of the population of black hole X-ray binaries via $RXTE$ data. The fractional rms of type C QPOs, which are computed in the energy band 2-26 keV, was found to statistically decrease with the frequency of LFQPOs for low-inclination sources, while the fractional rms-frequency trend for high-inclination sources apparently deviates from the low-inclination sources. 
More specifically, the derived fractional rms increases to the maximum at $\sim$2 Hz and then reduces beyond this frequency, with the average value being higher than that for low inclination. Furthermore, it should be noted that, in their work, there were spreads in the fractional rms around the average values for both low and high inclinations.

\begin{figure}
\includegraphics[height=13.0cm,width=21.0cm,trim=0.0cm 0.0cm 0.0cm
 0.0cm,clip=true]
{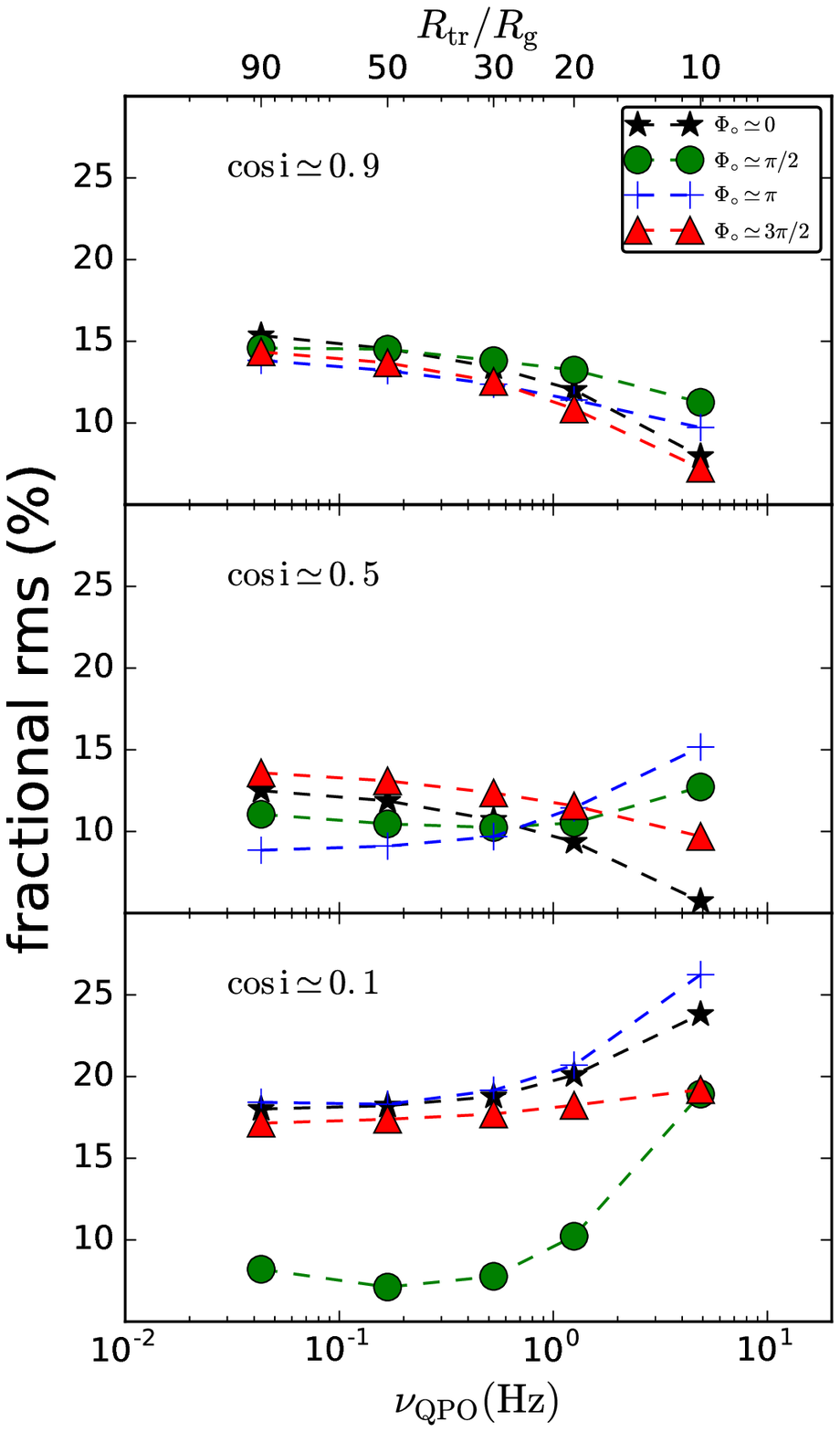}
\vspace{-1.5cm}
\caption{
Fractional rms as a function of the QPO centroid frequency for the half-opening angle of the flow $\theta_{\rm o} = 15^{\circ}$ in the energy band 2-26 keV. The truncation radius $R_{\rm tr}$ = 10, and the black hole spin $a=0.3$. The QPO frequency is computed based on Equation (2) of Ingram et al. (2009) as a function of truncation radius, which is labeled in the upper X-axis.
The top, middle, and bottom panels correspond to the sinusoidal value of viewing angle $\cos i$ = 0.9, 0.5, and 0.1, respectively. In each panel, the different point types correspond to different viewer azimuthal angles.
\label{rms_nu_phi}}
\end{figure}

In order to compare with the observational properties, we compute the fractional rms of LFQPOs from our simulation in the same energy band. The trend of the fractional rms with the frequency of LFQPOs is plotted in Fig. \ref{rms_nu_phi} for the different viewer azimuths, where the QPO frequency is computed based on Eq(2) of Ingram et al. (2009) as a function of truncation radius.
The top, middle, and bottom panels
correspond to low, moderate, and high inclination, i.e., $cos i \simeq$ 0.9, 0.5, and 0.1, respectively.
It can be seen that the fractional rms for low inclinations (top panel) decreases with QPO frequency. 
And in the case of low inclinations, the difference in the fractional rms spectrum due to different viewer azimuths is slight.
As the inclination increases (in the bottom panels), the fractional rms apparently differs from the one for low inclinations and the amplitude becomes larger, especially
for higher frequency. More importantly, here we find that, for moderate and high inclinations, the fractional rms shows the variation for different viewer azimuths, at a given QPO frequency. This may naturally explain the observed spread in the fractional rms in Motta et al. (2015), provided that the viewer azimuths of the objects in their paper perhaps are different. 

These results are consistent with the observational properties above, which support the hypothesis that type C LFQPOs may originate from the Lense-Thirring precession of the inner flow. As the source evolves from hard state to soft state, the truncation radius decreases, which will result in the evolution of both QPO frequency and variability amplitude, predicted by the precession model. On one hand, the frequency of LFQPOs increases with the truncation radius (Ingram et al. 2009). On the other hand, the variability amplitude depends on the source inclination and spectral state. At low inclinations, as discussed in previous sections, the variability mainly arises from the variation of the projected area due to the wobbling inner flow. Therefore, as the truncation radius decreases (evolving to soft state), i.e., QPO frequency increasing, the inner flow  becomes more compact and the wobbling effect will gradually weaken, so that the fractional rms amplitude correspondingly decreases. At high inclinations, the self-occultation between the accretion disk and torus plays an increasingly important role in the fractional rms amplitude, which then results in the deviation from the rms-frequency trend for low inclinations. 

\section{Summary and Conclusion}
In the geometry of the truncated disk, if the inner hot flow misaligns with the black hole spin, it will undergo the Lense-Thirring precession due to the frame-dragging effect. During the precession phase, the combination of the geometrical and relativistic effects contributes to the modulation of the X-ray flux: (1) the varying occultation of the truncated disk by the torus gives rise to the disk emission variability; (2) the varying optical depth along the line of sight, the heating-to-cooling ratio of the plasma, self-occultation, and the light-bending effect contribute to the modulation of the torus emission, which arises from upscattering the soft photons from the outer truncated disk; (3) the tilted flow under the precession will illuminate different azimuths of the disk, resulting in the blueshift/redshift modulation because of the rotation of the disk. 
      
We develop the Monte Carlo code to simulate the modulation of the relativistic X-ray spectrum by the Lense-Thirring precession of the truncated geometry, taking light bending and gravitational redshift into account. Based on these simulation results, we are able to investigate the evolution of the energy-dependent QPO variabilities during the spectral transition from hard to soft state, which is implemented by the decrease of the disk truncation radius. 
The main conclusions are as follows:
\begin{enumerate}
\item As the source in the outburst evolves from hard to soft state, the disk emission becomes variable with the fractional rms of roughly a few percent.
\item The fractional rms of the Comptonization increases with photon energy. Yet in the soft state, where the precession flow is compact, the fractional rms spectra are roughly constant across the energy.
\item The reflection variability amplitude increases with energy, because the illumination by soft photons can be relatively easily seen by both approaching and receding regions of the disk, so that the effect of the blueshift/redshift modulation is less than the illumination by hard photons.
\item The fractional rms computed in the energy range 2-26 keV is found to decrease with the QPO frequency at low inclination, while the fractional rms is enhanced to large amplitude at high inclination owing to the increasing role of self-occultation between the torus and accretion disk. These simulated fractional rms-frequency trends are consistent with the observational variability properties of type C LFQPO (Motta et al. 2015), supporting the hypothesis that type C LFQPO may originate from the Lense-Thirring precession of the inner flow.
 
\item It is plausible that the height (or the half-opening angle) of precessing flow is large in the hard state because of high temperature but decreases as the source evolves to soft state because of the increase in the cooling rate by the disk soft photon. Hence, the fractional rms amplitude could be up to $\sim 40\%$ in the hard state, while roughly a few percent in the soft state, which is consistent with the observation.
\item The fractional rms spectrum of the total radiation is weakly dependent on the spin, while the fractional variability amplitude of the reflection emission increases with photon energy, and the reflection for low spin is more variable than the one for high spin. 
\end{enumerate}

In our forthcoming papers, we will investigate the variability properties of the iron emission line due to the Lense-Thirring precession of the accretion flow (Ingram \& Done 2012), by the use of our Monte Carlo code developed in this work. Furthermore, we will also study both energy-dependent phase lag at both the fundamental and the first overtone and phase lag between two broad energy bands as a function of QPO frequency. The predictions will then be compared with the observational results (e.g, Ingram et al. 2015, 2016, 2017; van den Eijnden et al. 2017).

\section{Acknowledgments} 
We thank Barbara De Marco and Agata R\' o\. za\' nska for discussion and comments.
This research was supported by Polish National Science Centre grant 
nos. 2015/17/B/ST9/03422 and 2015/18/M/ST9/00541 and by Ministry of Science
and Higher Education grant W30/7.PR/2013. This research was supported by ``the Fundamental Research Funds for the Central Universities.'' It received funding from the 
European Union Seventh Framework Programme (FP7/2007-2013) under 
grant agreement no. 312789, Polish National Science Center grant Polonez 2016/21/P/ST9/04025, the National Program on Key Research and Development Project (grant no. 2016YFA0400803), and the Chinese National Science Foundation (grant no. 11622326).
This research was partly financed by grant DEC-2011/03/B/ST9/03459 from 
the Polish National Science Centre.
This work made use of the High Performance Computing Resource in the Core Facility for Advanced Research Computing at Shanghai Astronomical Observatory.

{}

\appendix
\section{Monte Carlo simulation of disk seed photon}
The flux in the local frame (comoving frame) at a given radius $r$ for a Novikov-Thorne accretion disk is denoted as $F(r)$, the formulae of which are given in Page \& Thorne(1974). The flux in the Boyer-Lindquist coordinate, $\widetilde{F}(r)$, can be transformed from the local flux (Kulkarni 2011):
\begin{equation}
\widetilde{F}(r) = -u_t\cdot F(r)
\end{equation}
where $u_t$ is the $t$-component of the fluid four-velocity $u^{\mu}$.
Therefore, at a given radius $r$, the effective temperature of the Novikov-Thorne accretion disk can be derived:
\begin{equation}
T_{\rm eff}(r) = (\widetilde{F}(r)/\sigma)^{1/4} = (-u_t\cdot F(r)/\sigma)^{1/4}
\end{equation}
Then, the intensity of the photon number
\begin{equation}
N_{\nu}(r) = B_{\nu}(T_{\rm eff})/h\nu
\end{equation} 
where $B_{\nu}$ is the blackbody radiation intensity. Integrating over frequency, the number of photons emitted from the disk at radius $r$ is
\begin{equation}
N_{\rm T}(r) = \int_{0}^{\infty}N_{\nu}(r) d\nu
\end{equation} 
Therefore, we can define the cumulative distribution function of photon number:
\begin{equation}
CDF(r) = \frac{\int_{r_{in}}^{r}N_{\rm T}(r)\cdot r\cdot dr}{\int_{r_{in}}^{r_{out}}N_{\rm T}(r)\cdot r\cdot dr}
\end{equation} 
where $r_{\rm in}$ and $r_{\rm out}$ are the inner and outer radius of the disk, respectively. It can be easily seen that CDF($r$) ranges from 0 to 1, monotonically increasing with radius $r$. Consequently, it is more useful to build the inverse cumulative distribution function iCDF, so that random number ``$n$'' between 0 and 1 is used in iCDF($n$) to uniquely determine the emitting radius of the Novikov-Thorne accretion disk and the corresponding temperature. 
Then, photon energy will be randomized based on the Planck function.  

The remaining information of the disk soft photons is the azimuth of the emitting position and the direction of the motion.
The azimuth of the emitting position is randomly generated between 0 and $2\pi$. The directions of the motion are specified by two angles in the local frame, i.e., the azimuthal angle $\phi$ and the polar angle $\theta$ with respect to the normal of the disk. The azimuthal angle $\phi$ is randomly generated between 0 and $2\pi$, while the polar angle $\theta$ is randomly generated under the condition that the number of photons per second emitted into equal portions of solid angle depends on the polar angle. More specifically, at a given radius $r$ and azimuthal angle $\phi$ of the photon's motion, the number of photons per second emitted into equal portions of solid angle is
\begin{equation}
N(\theta\mid\phi, r) = N_{\rm T}(r)\cdot \cos\theta \cdot dA\cdot d\Omega
\end{equation} 
As the solid angle $d\Omega = \sin\theta \cdot d\theta \cdot d\phi$, we have
\begin{equation}
N(\theta\mid\phi, r) = N_{\rm T}(r)\cdot \cos\theta \cdot \sin\theta \cdot d\theta \cdot dA\cdot d\phi
\end{equation}
Therefore, we can construct the cumulative distribution function of photon number as a function of the polar angle:
\begin{equation}
CDF(\theta) = \frac{\int_{0}^{\theta}N(\theta\mid\phi, r)d\theta}{\int_{0}^{\pi/2}N(\theta\mid\phi, r)d\theta}
\end{equation} 
The CDF$(\theta)$ ranges from 0 to 1, monotonically increasing with the polar angle $\theta$. Consequently, we build the inverse cumulative distribution function iCDF$(n)$, so that random number ``$n$'' between 0 and 1 is used in iCDF$(n)$ to uniquely determine the polar angle. Note that the disk photons with the polar angle ranging from $\pi/2$ to $\pi$ are also included in the code.

\end{document}